\documentclass[lettersize,journal]{IEEEtran}
\usepackage{cite}
\usepackage{amsmath,amssymb,amsfonts}
\usepackage{algorithmic}
\usepackage{graphicx}
\usepackage{textcomp}
\usepackage{xcolor}
\usepackage[ruled,vlined,linesnumbered,commentsnumbered]{algorithm2e}
\usepackage{booktabs}
\usepackage{url}
\usepackage{graphicx}
\usepackage{colortbl}
\usepackage{multicol}
\usepackage{multirow}
\usepackage{arydshln}
\usepackage{amsmath}
\usepackage{pifont}
\usepackage{booktabs}
\usepackage{bm}
\usepackage{wrapfig}
\usepackage{algorithmic}
\usepackage{threeparttable}
\usepackage{hyperref}
\usepackage{enumitem}
\usepackage{subcaption}

\newtheorem{lemma}{Lemma}
\newtheorem{proposition}{Proposition}
\newcommand{\method}{UFO}
\newcommand{\tabincell}[2]{\begin{tabular}{@{}#1@{}}#2\end{tabular}}
\begin{document}

\title{
    \huge
    \textbf{
    \method: Unlocking Ultra-Efficient Quantized Private Inference with Protocol and Algorithm Co-Optimization
    }
}

\author{
  Wenxuan Zeng$^{12}$, Chao Yang$^{12}$, Tianshi Xu$^{13}$, Bo Zhang$^4$, Changrui Ren$^4$, Jin Dong$^4$, Meng Li$^{13\dag}$ \\
  \vspace{5pt}
  \textit{$^1$Institute for Artificial Intelligence, Peking University, China} \\
  \textit{$^2$School of Software and Microelectronics, Peking University, China} \\
  \textit{$^3$School of Integrated Circuits, Peking University, China} \\
  \textit{$^4$Beijing Academy of Blockchain and Edge Computing} \\
  \thanks{
    $\dag$ Corresponding author: meng.li@pku.edu.cn
  }
}



\maketitle

\begin{abstract}

Private convolutional neural network (CNN) inference based on secure two-party computation (2PC) suffers from high 
communication and latency overhead, especially from convolution layers.
In this paper, we propose \method, a quantized 2PC inference framework that jointly optimizes the 2PC protocols and quantization algorithm.
\method~features a novel 2PC protocol that systematically combines the efficient Winograd convolution algorithm with quantization
to improve
inference efficiency.
However, we observe that naively combining quantization and Winograd convolution faces the following challenges:
\textit{\underline{1)} From the inference perspective,} Winograd transformations introduce extensive additions and require frequent bit width conversions to avoid inference overflow, leading to non-negligible communication overhead;
\textit{\underline{2)} From the training perspective,} Winograd transformations introduce weight outliers that make quantization-aware training (QAT) difficult, resulting in inferior model accuracy.
To address these challenges, we co-optimize both protocol and algorithm.
\textit{\underline{1)} At the protocol level,} we propose a series of graph-level optimizations for
2PC inference to minimize the communication. 
\textit{\underline{2)} At the algorithm level,}
we develop a mixed-precision QAT algorithm based on layer sensitivity
to optimize model accuracy given communication constraints.
To accommodate the outliers, we further introduce a 2PC-friendly bit re-weighting algorithm to increase the representation range without explicitly increasing bit widths.
With extensive experiments, \method~demonstrates 11.7$\times$, 3.6$\times$, and 6.3$\times$
communication reduction with 1.29\%, 1.16\%, and 1.29\% higher accuracy compared
to state-of-the-art frameworks SiRNN, COINN, and CoPriv, respectively.

% \wx{No need to emphasize 2PC? MPC is OK}

% \begin{IEEEkeywords}
%     2PC-based private inference, quantized Winograd convolution protocol, bit re-weighting, mixed-precision quantization.
% \end{IEEEkeywords}

\end{abstract}    
\section{Introduction}
\label{sec:intro}

Private deep neural network (DNN) inference based on secure two-party computation (2PC) provides cryptographically strong data privacy protection
% and has attracted more and more attention in recent years 
\cite{rathee2020cryptflow2,rathee2021sirnn,mohassel2017secureml,mishra2020delphi,liu2017oblivious,shen2022abnn2,zeng2025towards}.
2PC inference frameworks target protecting the privacy of both model parameters held by the server and input data held by the client. 
By jointly executing a series of 2PC protocols, the client can learn the final inference results but nothing else on the model can be derived from the results.
Meanwhile, the server knows nothing about the client's input \cite{rathee2021sirnn,rathee2020cryptflow2,mohassel2017secureml,mishra2020delphi,jha2021deepreduce,lu2023squirrel,shen2022abnn2,mohassel2018aby3,liu2017oblivious}.

\begin{figure}[!tb]
    \centering
    \includegraphics[width=\linewidth]{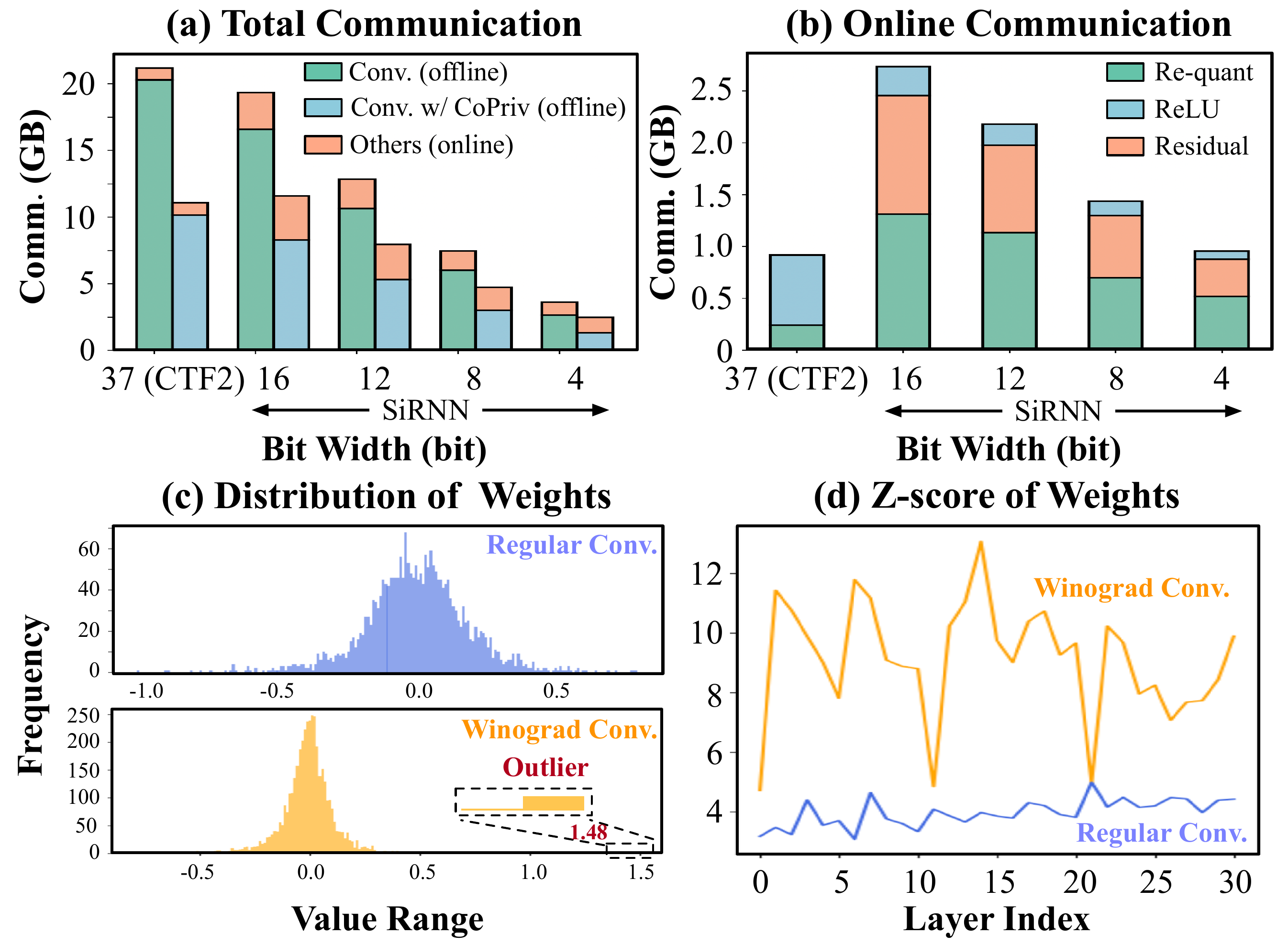}
    \caption{Motivating inspirations. (a) Total communication and (b) online communication breakdown on the ResNet-50 block profiled with
    CrypTFlow2 (CTF2) \cite{rathee2020cryptflow2} (uniform 37-bit) and SiRNN \cite{rathee2021sirnn} (low precision); 
    (c) weight distributions in regular and Winograd convolution;
    (d) z-score (defined as the \textit{ratio between max-average and standard deviation}) indicates weight outliers consistently exist after Winograd transformation across different layers.}
    \label{fig:intro_bar}
    % \vspace{-10pt}
\end{figure}

However, 2PC frameworks achieve high privacy protection at the cost of
orders of magnitude latency and communication overhead due to the massive interaction
between the server and client. 
As shown in Figure~\ref{fig:intro_bar}(a) and (b),
the total communication of a convolutional neural network (CNN)
is dominated by its convolution layers while the online 
communication is generated by non-linear functions, e.g., ReLU.
To improve communication efficiency, a series of works have proposed algorithm and protocol optimizations
% network and protocol optimizations have been proposed
\cite{zeng2023copriv,riazi2019xonn,hussain2021coinn,rathee2021sirnn,
shen2022abnn2}. 
Typically, CoPriv \cite{zeng2023copriv} proposes a Winograd-based
convolution protocol to reduce the communication-expensive multiplications
at the cost of more local additions. Although it achieves 2$\times$ communication
reduction, it still requires more than 2GB of communication for a single
ResNet-18 block. Recent works also leverage mixed-precision quantization for communication
reduction \cite{hussain2021coinn,rathee2021sirnn,
shen2022abnn2}. Although total communication reduces
consistently with the inference bit widths,
as shown in Figure~\ref{fig:intro_bar}(b), existing mixed-precision protocols suffer from
much higher online communication even for 4 bits
due to complex protocols like re-quantization and residual.

As the communication of 2PC frameworks scales with both the inference bit widths
and the number of multiplications, we propose to combine the 
Winograd convolution with low-precision quantization. However, we
observe a naive combination leads to limited improvement. On one hand,
although the local additions in the Winograd transformations do not
introduce communication directly, they require higher inference bit widths to avoid inference overflow
and complicate the bit width conversion protocols. 
As a result, a naive
combination only reduces $\sim$20\% total communication with even more online 
communication compared to not using Winograd convolution as shown in Figure~\ref{fig:intro_bar}(a).
On the other hand, the
Winograd transformations also distort the model weight distribution such that
introduce more outliers that make quantization harder, as shown in Figure \ref{fig:intro_bar}(c) and (d). 
Hence, training a quantized model with outliers inevitably results in inferior model accuracy.

% \section{Contributions}
In this paper, we propose an efficient 2PC framework named 
\method~to solve the aforementioned challenges. \method~systematically combines Winograd convolution with
quantization and features a series of protocol and algorithm
optimizations. Our contributions
can be summarized below:
\begin{itemize}
% [leftmargin=*,itemsep=2pt,topsep=2pt]
    \item \textbf{At the protocol level.} We observe the communication of 2PC inference scales with both
    the bit widths and the number of multiplications. Hence, we propose
    \method, a novel convolution protocol that systematically combines Winograd convolution and mixed-precision
    quantization for efficient inference.
    \method~features a series of graph-level optimizations
    to reduce online communication.
    \item \textbf{At the algorithm level.} We propose a 2PC-aware mixed-precision Winograd quantization-aware training (QAT) algorithm and further develop a 2PC-friendly bit re-weighting algorithm to
    handle the outliers introduced by the Winograd transformations
    and improve the accuracy.
\end{itemize}

\textbf{Evaluation results.} Extensive experiments demonstrate that \method~achieves 11.7$\times$, 3.6$\times$, and 6.3$\times$ communication reduction with 1.29\%, 1.16\%, and 1.29\% higher accuracy compared to state-of-the-art frameworks SiRNN, COINN, and CoPriv, respectively.

\section{Preliminaries and Backgrounds}
\label{sec:pre}

\begin{figure}[!tb]
    \centering
    \includegraphics[width=\linewidth]{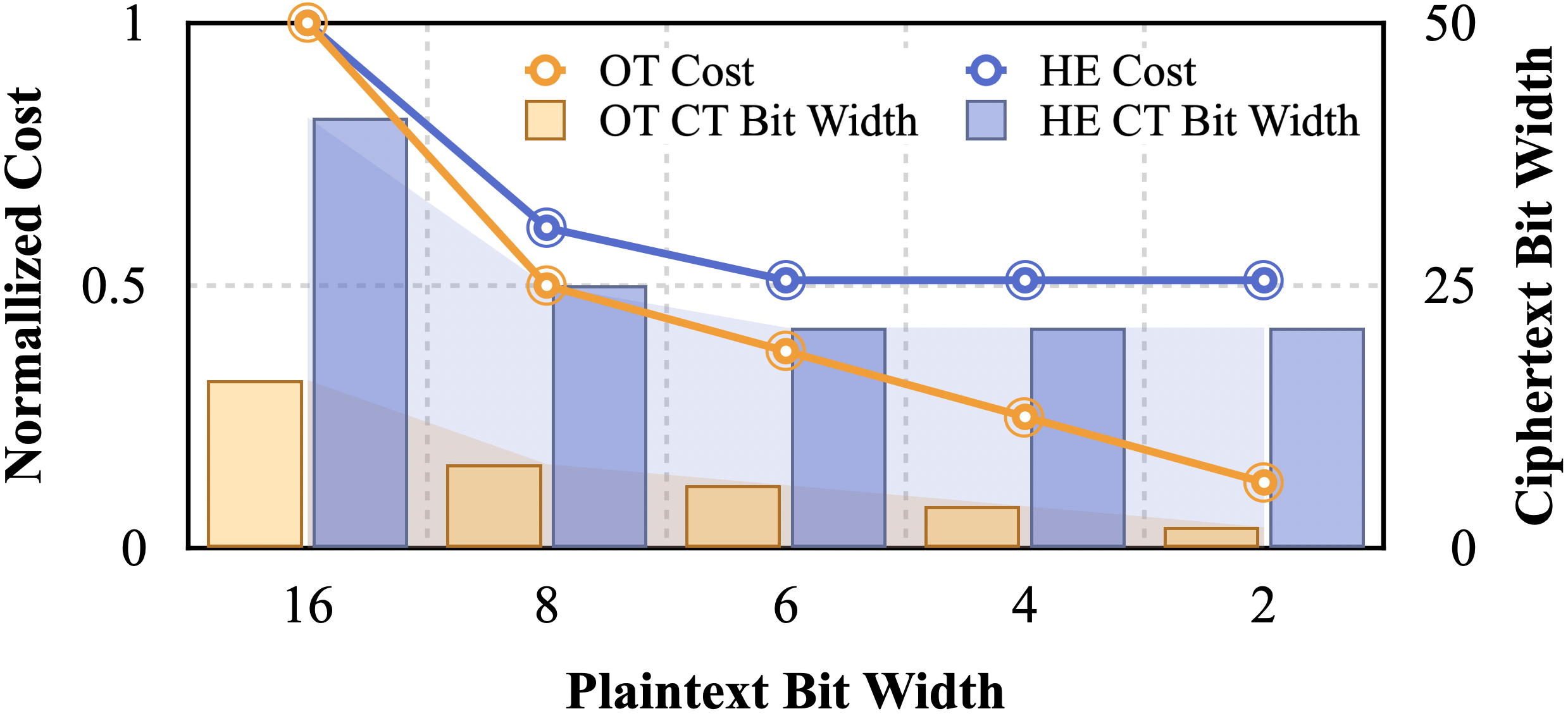}
    \caption{Comparison between OT and HE with quantization. HE requires much higher ciphertext (CT) bit widths and cannot utilize low precision, e.g., 4 bits.}
    \label{fig:compare_mpc_he_quant}
\end{figure}

\subsection{Threat Model}

\method~is a 2PC-based private inference framework that involves a client and a server.
The server holds the private model weights and the client holds the private input data.
Following \cite{rathee2021sirnn,rathee2020cryptflow2,mohassel2017secureml,mishra2020delphi,cho2022selective,jha2021deepreduce,kundu2023learning},
we assume the DNN architecture (including the number of layers and the operator type, shape, 
and bit widths) are known by two parties.
At the end of the protocol execution, the client learns the inference result and the two parties know nothing else
about each other's input.
Consistent with prior works, \method~adopts a \textit{honest-but-curious} security model\footnote{Malicious security model is also an important research direction but is not the focus
of this paper.} where both parties follow the protocol specifications but also attempt to learn more from the information than allowed.
We also assume no trusted third party exists 
% so the helper data needs to be generated by the client and server 
\cite{rathee2020cryptflow2,rathee2021sirnn,mishra2020delphi}.
% \method~is built upon underlying OT primitives, which are proven to be secure in the \textit{honest-but-curious}
% security model \cite{rathee2020cryptflow2,rathee2021sirnn}. 
% Utilizing quantized models and Winograd convolution does not affect OT primitives in any way.
We also elaborate on the security and correctness analysis of \method~in Section \ref{sec:security}.
% \subsection{Underlying Protocols and Notations}
For clarity, we summarize the protocols used in this paper in Table \ref{tab:protocols} and notations in Table \ref{tab:notation}.

\begin{figure*}[!tb]
    \centering
    \includegraphics[width=\linewidth]{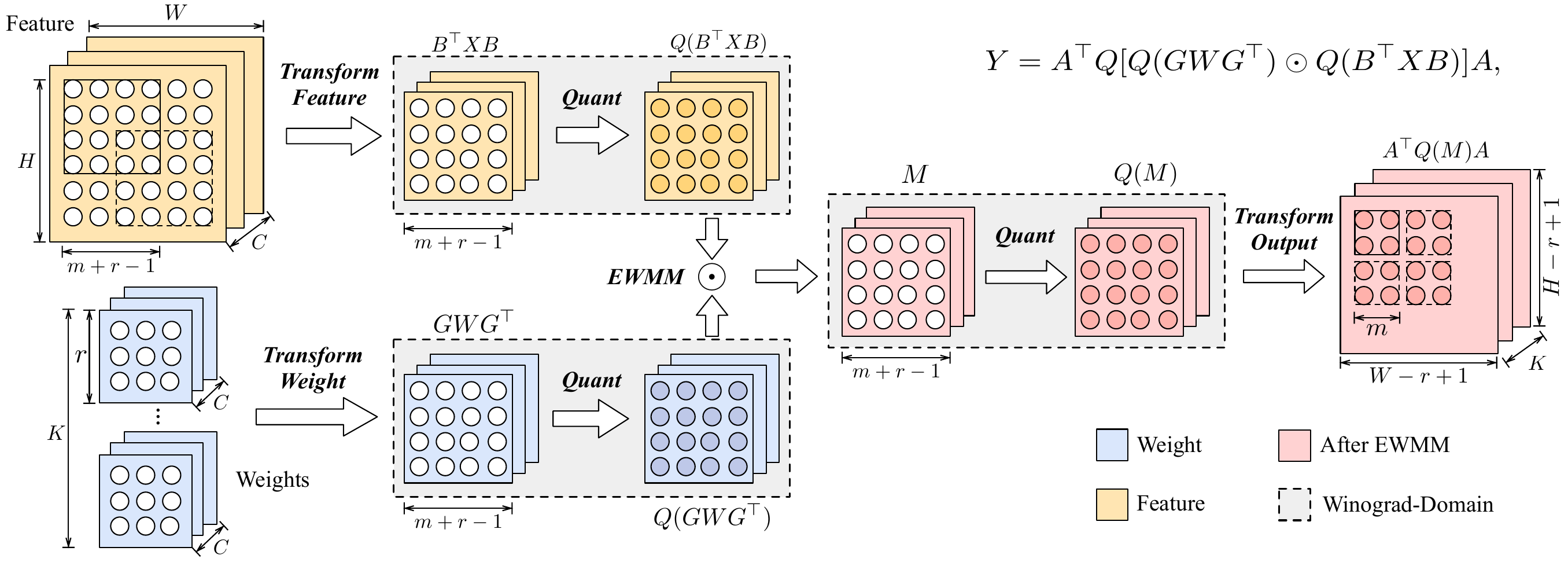}
    \caption{Overall pipeline of Winograd convolution with low-precision quantization.}
    \label{fig:winograd_qat}
\end{figure*}

% Following prior work, we assume the server and client are semi-honest
% adversaries $\mathcal{A}$. 
% Formally, our security analysis follows the widely accepted simulation paradigm targeting static semi-honest probabilistic polynomial time (PPT) adversaries. The core principle is to prove that the view of an adversary $\mathcal{A}$ in a real execution is indistinguishable from a simulated view in an ideal execution, where the computation is handled by a trusted functionality. Specifically, security holds if a simulator $\mathcal{S}$ can be constructed in the ideal setting such that no environment $\mathcal{Z}$ can differentiate between the real and ideal interactions. Furthermore, we employ the hybrid model for modularity, treating sub-protocols as calls to ideal functionalities F within our analysis.

\begin{table*}[!tb]
    \centering
    \caption{Underlying protocols used in this paper.}
    \label{tab:protocols}
    \resizebox{\linewidth}{!}{
    \begin{tabular}{l|l|l}
    % \bottomrule
    % \textbf{Protocol}     &  \textbf{Description} &  \textbf{Comm. Complexity} \\
    % \hline
    % Bit Extension     &  Extend tensor with $l_1$-bit to $l_2$-bit & $O(\lambda (l_1+1))$ \\
    % \hline
    % Truncation   &  \tabincell{c}{Truncate (right shift) tensor with \\$l_1$-bit by $l_2$-bit}  &  $O(\lambda (l_1+3))$  \\
    % \hline
    % Re-quantization  &   \tabincell{c}{Re-quantize tensor with $l_1$-bit and $s_1$ \\ scale to tensor with $l_2$-bit and $s_2$ scale} & \tabincell{c}{Combination of extension \\and truncation (refer to \cite{xu2024privquant})}  \\
    \bottomrule
    \textbf{Underlying Protocol}  &  \textbf{Description} & \textbf{Communication Complexity}   \\
    \hline
    $\langle z \rangle^{(l_2)}=\Pi_{\mathrm{Ext}}^{l_1, l_2}(\langle x \rangle^{(l_1)})$  & \tabincell{c}{Extend $l_1$-bit $x$ to $l_2$-bit $z$ such that $z^{(l_2)} = x^{(l_1)}$} &   $O(\lambda (l_1+1)+13l_1+l_2)$ \\
    $\langle z \rangle^{(l_1)}=\Pi_{\mathrm{Trunc}}^{l_1,l_2}(\langle x \rangle^{(l_1)})$ &  \tabincell{c}{Truncate (right shift) $l_1$-bit $x$ by $l_2$-bit such that $z^{(l_1)}=x ^{(l_1)} \gg l_2$} & $O(\lambda (l_1+3)+15l_1+l_2+20)$ \\
    $\langle z \rangle^{(l_1-l_2)}=\Pi_{\mathrm{TR}}^{l_1,l_2}(\langle x \rangle^{(l_1)})$ & \tabincell{c}{Truncate $l_1$-bit $x$ by $l_2$-bit and discard the high $l_2$-bit such that $z^{(l_1-l_2)}=x^{(l_1)} \gg l_2$} & $O(\lambda (l_2+1)+13l_2+l_1)$ \\
    $\langle z\rangle^{(l_2)}=\Pi_{\mathrm{Requant}}^{l_1,s_1,l_2,s_2}(\langle x \rangle^{(l_1)})$  &  \tabincell{c}{Re-quantize $x$ with $l_1$-bit and $s_1$ scale to $z$ with $l_2$-bit and $s_2$ scale} & 
    \tabincell{c}{Combination of $\Pi_{\mathrm{Ext}},\Pi_{\mathrm{Trunc}},\Pi_{\mathrm{TR}}$} \\
    \bottomrule
    \end{tabular}
    }
\end{table*}

\begin{table}[!tb]
    \centering
    \caption{Notations used in this paper. 
    }
    \label{tab:notation}
    \resizebox{\linewidth}{!}{
    \begin{tabular}{c|c}
    \bottomrule
    \textbf{Notation}     & \textbf{Description} \\
    \hline
    $\lambda$ & Security parameter that measures the attack hardness \\
    \hline
    $l_w, l_a, l_{acc}, l_{res}, l_{add}$ & \tabincell{c}{Bit width of weight, activation, accumulation, \\residual, and addition} \\
    \hline
    $l_{ft\_ext}, l_{out\_ext}$  & \tabincell{c}{Bit width of feature and output transformation \\after bit extension} \\
    \hline
    $s_a, s_{acc}, s_{res}, s_{add}$ &  \tabincell{c}{Scale of activation, accumulation, residual, \\and addition} \\
    \hline
    $x^{(l)}, \langle x \rangle^{(l)}$ & $l$-bit integer $x$ and $l$-bit secret share \\
    \hline
    % $H, W, C, K$  & \tabincell{c}{Height and width of output feature, \\and number of input and output channel} \\
    % \hline
    $A, B, G$  & Winograd transformation matrices \\
    \hline
    W$l_w$A$l_a$ & $l_w$-bit weight and $l_a$-bit activation   \\
    \bottomrule
    \end{tabular}
    }
\end{table}

\subsection{Private Inference}
Existing 2PC inference can be divided into OT-based and homomorphic encryption (HE)-based.
HE-based methods \cite{huang2022cheetah,hao2022iron,xu2023falcon,pang2023bolt,lu2023bumblebee,xu2025breaking,juvekar2018gazelle,he2024rhombus,ju2024neujeans,moon2025thor,yu2024flexhe} use HE to compute linear layers with lower communication compared to OT-based methods at the cost of more computation for both the server and client. 
Hence, they have different applicable scenarios \cite{hussain2021coinn,zeng2023copriv,xu2024privquant}. 
% For example, for resource-constrained clients, HE may not be applicable since it involves repetitive
% encoding, encryption, and decryption on the client side \cite{krieger2023aloha,gupta2022memfhe}.
% Hence, we focus on optimizing OT-based inference to reduce the communication in this work

% \subsubsection{\textbf{Protocol Selection}}
In this paper, we focus on OT-based methods instead of HE-based
methods.
On the one hand, HE usually requires the client to have a high computing capability for encryption and decryption.
On the other hand, HE is not friendly to quantization.
As shown in Figure \ref{fig:compare_mpc_he_quant}, there are two main reasons: 
1) HE requires significantly higher ciphertext bit widths to ensure decryption correctness~\cite{fan2012somewhat};
2) HE cannot utilize low-precision quantization (e.g., 4-bit) because suitable prime moduli are unavailable at such low bit widths~\cite{xu2024hequant}.

We leave detailed related work in Section \ref{sec:related}, and we compare \method~with previous works qualitatively in Table \ref{tab:compare_exist}. \method~jointly optimizes both protocol and algorithm to simultaneously reduce the number of multiplication and communication per multiplication for efficient 2PC inference.

% \ml{For all these references, are they all for OT-based method?}\zwx{Almost expect Iron \cite{hao2022iron}, CryptoGCN \cite{ran2022cryptogcn}, and LinGCN \cite{peng2023lingcn}}

% We leave the detailed review of existing works to Appendix \ref{related}.

\subsection{Secret Sharing and OT-based Linear Layer}
\label{sec:arss}

\begin{figure}[!tb]
    \centering
    \includegraphics[width=\linewidth]{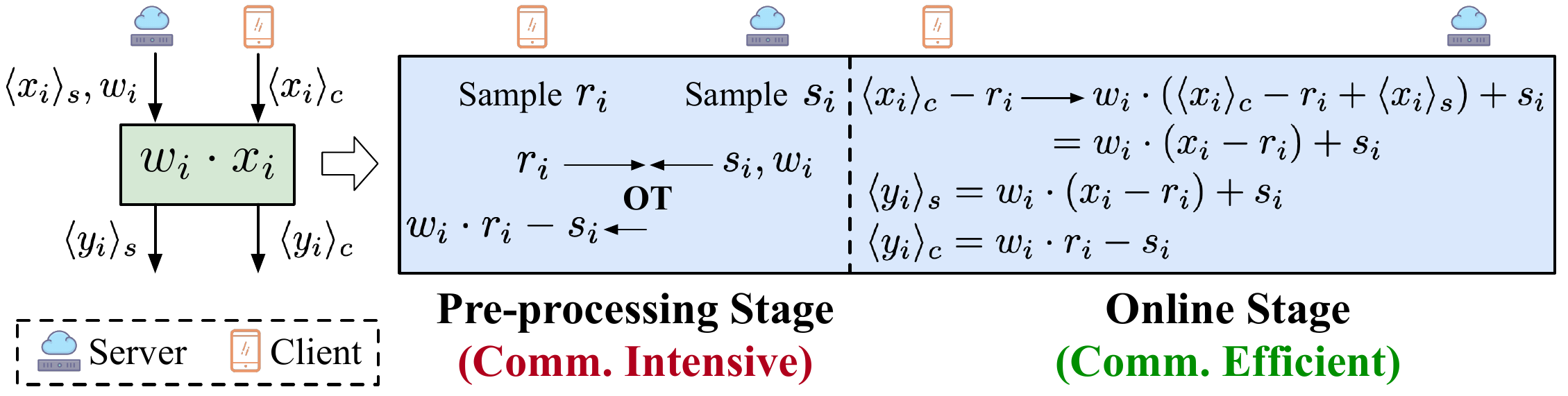}
    \caption{Protocol of OT-based linear layer, including a pre-processing stage and an online stage to process client's input.}
    \label{fig:ot_matmul}
\end{figure}

Figure~\ref{fig:ot_matmul} demonstrates the protocol of 2PC inference based on OT \cite{naor2001efficient,mishra2020delphi,rathee2020cryptflow2}. With arithmetic secret sharing,
each intermediate activation tensor in the $i$-th layer, e.g., $x_i$, is additively shared where the server holds $\langle x_i\rangle_s$
and the client holds $\langle x_i\rangle_c$ such that $x_i = \langle x_i\rangle_s+\langle x_i\rangle_c$ \cite{mohassel2017secureml}.
To generate the result $y_i$ of a linear layer, e.g., general matrix multiplication (GEMM),
the server and client jointly execute computation at the pre-processing (offline) stage and online stage \cite{mishra2020delphi}. 
In the pre-processing stage, the client and server first sample random $r_i$ and $s_i$, respectively.
Then, $\langle y_i\rangle_c=w_i\cdot r_i-s_i$ can be computed with a single OT if $r_i \in \{0, 1\}$ for 1 bit.
% With the vector optimization \cite{hussain2021coinn}, one OT can be extended to compute $w_i \cdot \bm{r_i} - \bm{s_i}$,
% where $\bm{r_i}$ and $\bm{s_i}$ are both vectors.
When $w_i$ has $l_w$ bits, we can repeat the OT protocol $l_w$ times by computing $w_i^{(b)} \cdot r_i-s_i$ each time,
where $w_i^{(b)}$ denotes the $b$-th bit of $w_i$. The final results can then be acquired by shifting and adding the partial results together.
Compared with the pre-processing stage, the online stage only requires very little communication to obtain $\langle y_i\rangle_s=w_i\cdot(x_i-r_i)+s_i$.

\subsection{Model Quantization}
Quantization converts a floating-point number into integers for better training and inference efficiency \cite{krishnamoorthi2018quantizing,lin2025qserve,wei2025lightmamba}, which is particularly beneficial for deployment on resource-constrained devices.
Specifically, a floating point number $x_f$ can be approximated by an $l_x$-bit integer $x_q$ and a scale $s_x$ 
through quantization as $x_q / s_x$,
% \footnote{We consider symmetric quantization without zero shiftand force $s_x$ to be power of 2 to reduce communication following \cite{rathee2021sirnn}.},
where
\begin{equation*}
    x_q = \max(-2^{l_x-1}, \min(2^{l_x-1}-1, \mathrm{round}(s_x x_f))).
\end{equation*}

The multiplication of two floating point numbers 
$x_f$ and $w_f$, denoted as $y_f$, can be approximately computed as $x_q w_q / (s_w s_x)$,
which is a quantized number with $(l_x + l_w)$-bit and $s_w s_x$ scale.
Then, $y_f$ usually needs to be re-quantized to $y_q$ with $l_y$-bit and $s_y$ scale as follows:
\begin{equation*}
    y_q = \max(-2^{b_y-1}, \min(2^{b_y-1}-1, \mathrm{round}(\frac{s_y}{s_w s_x} w_q x_q))).
\end{equation*}

For the addition, e.g., residual connection of two quantized numbers $x_q$ and $y_q$, directly adding them together leads to incorrect results. Instead, the scales and the bit widths of $x_q$ and $y_q$ need
to be aligned first.

\subsection{Winograd Convolution}
Winograd convolution \cite{lavin2016fast} is an efficient convolution algorithm that minimizes the number of multiplications.
% Winograd transformation $F(m \times m, r \times r)$, where the sizes of the output tile and weight are
% $m \times m$ and $r \times r$, respectively, can be formulated as
% \begin{equation*}
Winograd convolution can be formulated as
\begin{equation}
\label{eq:winograd}
    Y = W \circledast X = A^\top [(G W G^\top) \odot (B^\top X B)]A,
\end{equation}
where $\circledast$ denotes regular convolution and $\odot$ denotes element-wise matrix multiplication (EWMM). 
The overall pipeline is illustrated in Figure \ref{fig:winograd_qat}, and we also show where to insert the quantization operation in Winograd convolution.
We follow \cite{zeng2023copriv} to convert EWMM into GEMM with batch optimization.
$A$, $B$, and $G$ are constant transformation matrices
that are independent of the $W$ and $X$ and can be pre-computed 
% based on the tile size
based on the size of the convolution kernel and output tile $m$ and $r$
\cite{lavin2016fast}.
Weight transformation $G W G^\top$ is performed offline
while feature transformation $B^\top X B$ and output transformation $A^\top [\cdot] A$ are performed online.
The number of multiplications can be reduced from $m^2r^2$ to $(m+r-1)^2$ at the cost of many more additions introduced by the feature and output transformation.
% The number of multiplications can be reduced at the cost of more additions introduced by the feature and output transformation.
\textbf{For 2PC inference, since transformation matrices are constant and public, $B^\top X B$ and $A^\top [\cdot] A$ can be computed locally without any communication. However, Winograd-domain GEMM incurs 2PC communication costs.}
% \cite{zeng2023copriv} further formulates the EWMM into the form of general matrix multiplication for better communication efficiency.
% As a consequence, the communication complexity is reduced from $O(\lambda CKT(m+r-1)^2)$ to $O((m+r-1)^2CT(\lambda+K))$,
% where $C, K$, and $T$ denote the number of input channels, output channels, and the number of tiles, respectively.
% We refer interested readers to \cite{zeng2023copriv} for more details.

Considering the potential of quantization, some works combine Winograd and quantization \cite{barabasz2020quantaized,chikin2022channel,chen2024towards,mori2024wino,pan2025data}; however, they are not tailored for 2PC, resulting in suboptimal performance.

% \wx{Add a figure of Winograd workflow.}

\section{Motivations}
\label{sec:motivation}

In this section, we first theoretically analyze the communication complexity of the OT-based linear layer, and then highlight the key challenges when combining Winograd convolution with quantization,
which motivates \method.

\begin{figure}[!tb]
    \centering
    \includegraphics[width=\linewidth]{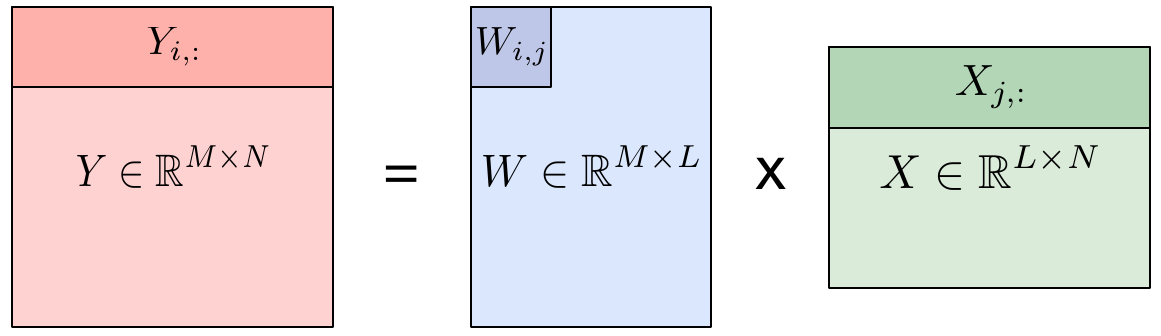}
    \caption{An illustration of GEMM $Y=WX$.}
    \label{fig:matmul}
\end{figure}

\subsubsection{\textbf{Motivation 1: the total communication of OT-based 2PC inference is determined by both the bit widths and the number of multiplications in linear layers}}
Consider $Y = WX$, 
where $W\in \mathbb{R}^{M\times L}$, $X\in \mathbb{R}^{L\times N}$ and $Y\in \mathbb{R}^{M\times N}$.
With one round of OT, we can compute $W_{i, j}^{(b)} \cdot X_{j, :}$ for the $b$-th bit of $W_{i, j}$ and $j$-th row of $X$.
% \ml{Need to unify with later sections.}.
Then, the $i$-th row of $Y$, denoted as $Y_{i,:}$, can be computed as
$$
    \label{eq:ot}
    Y_{i, :} = \sum_{j=0}^{L-1} \sum_{b=0}^{l_w-1} 2^b \cdot W_{i, j}^{(b)} \cdot X_{j, :},
$$
where $l_w$ denotes the bit width of $W$.
Hence, to compute $Y_{i, :}$, in total $O(l_wL)$ OTs are invoked.
% In each OT, as $X_{j, :}$ are transferred,
In each OT, the communication scales proportionally
with the vector length of $X_{j, :}$, i.e., $O(N l_x)$, where $l_x$ denotes the bit width of $X$.
The total communication of the GEMM thus becomes $O(MLN l_w l_x)$.
\textit{Hence, we observe the total communication of a GEMM scales with both the operands' bit widths,
i.e., $l_x$ and $l_w$, and the number of multiplications, i.e., $MLN$,
both of which impact the round of OTs and the communication per OT.}
Convolutions follow a similar relation as GEMM. Hence, combining Winograd convolution and quantization
is a promising solution to improve the communication efficiency of convolution layers.
For non-linear layers, e.g., ReLU,
communication scales proportionally with activation bit widths \cite{rathee2021sirnn}.

\begin{figure}
    \centering
    \includegraphics[width=\linewidth]{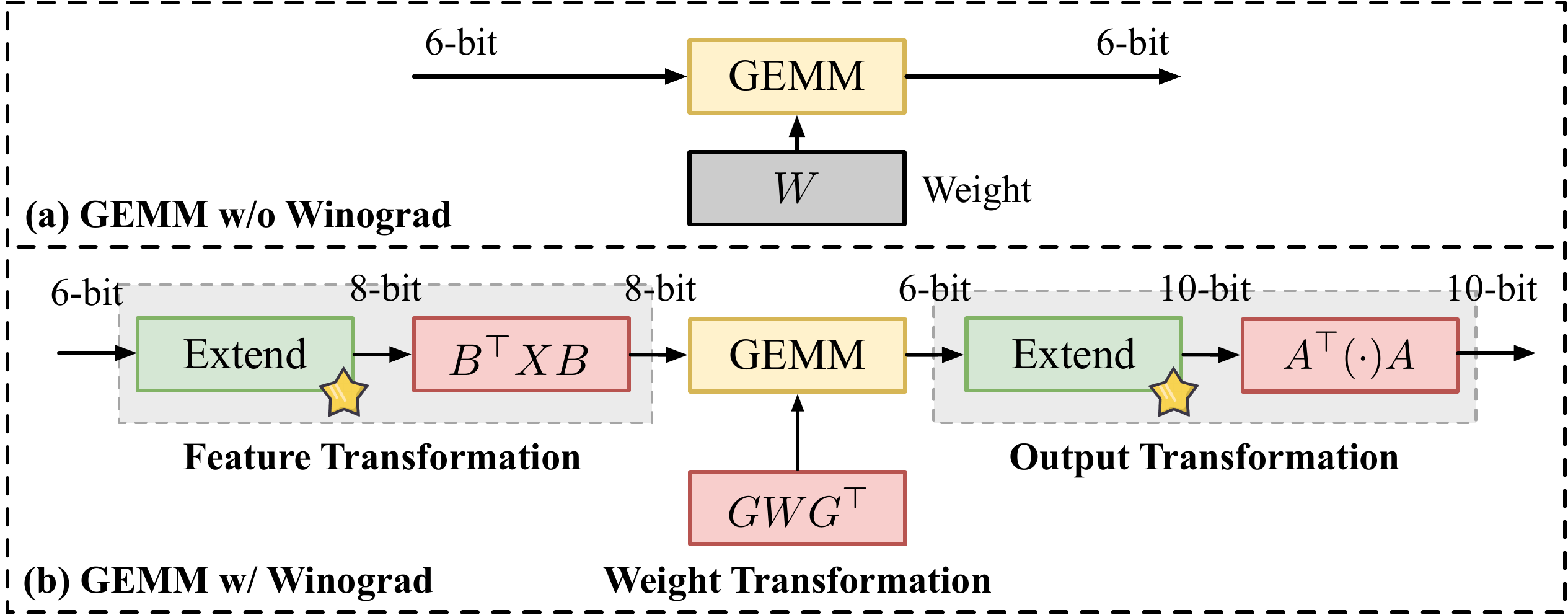}
    \caption{Bit width conversions, i.e., extension (yellow stars) are required to avoid inference overflow during Winograd transformations with quantized 2PC inference. There is a re-quantization in GEMM to output 6-bit activation.}
    \label{fig:compare_extend}
\end{figure}

\begin{figure}
    \centering
    \includegraphics[width=\linewidth]{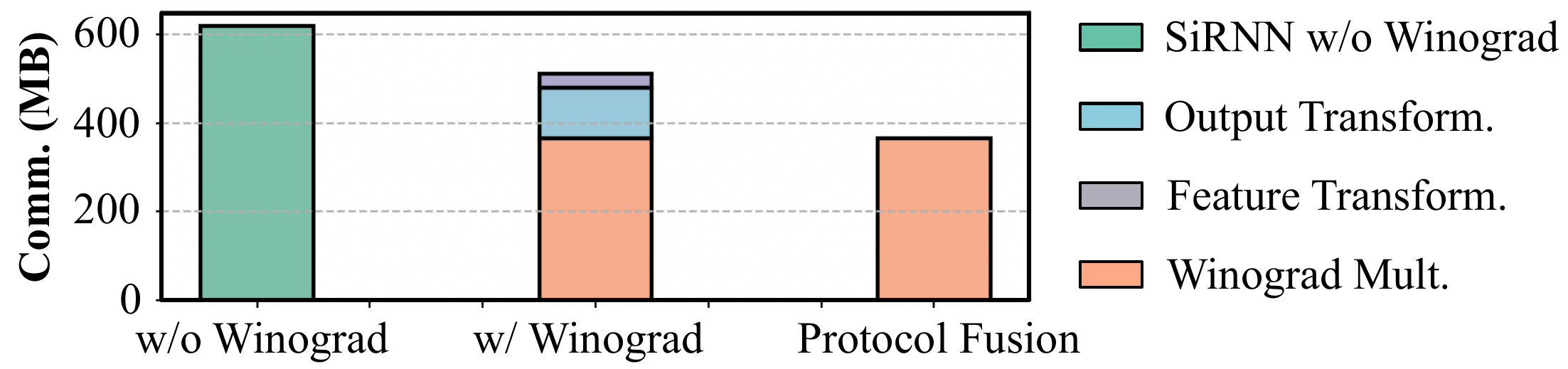}
    \caption{Communication breakdown on ResNet-32. After protocol fusion (introduced in Section \ref{sec:fusion}), the communication of both feature and output transformation are totally removed.}
    \label{fig:winograd_breakdown}
\end{figure}

\begin{figure}[!tb]
    \centering
    \includegraphics[width=\linewidth]{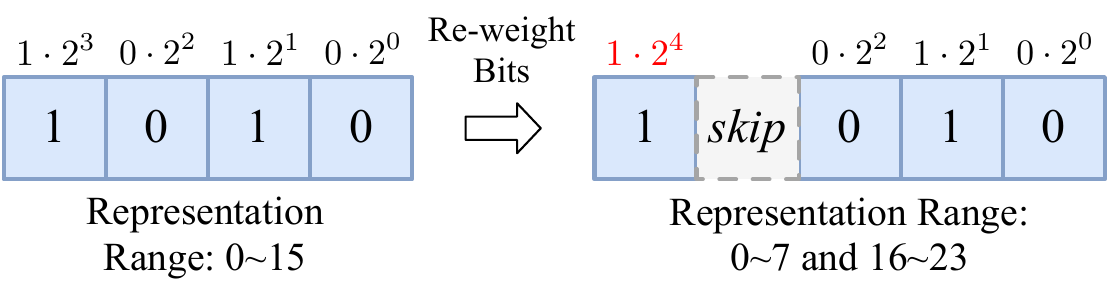}
    \caption{Example of bit re-weighting with adjusted representation range.}
    \label{fig:skip}
    % \vspace{-15pt}
\end{figure}

\begin{figure*}[!tb]
    \centering    
    \includegraphics[width=\linewidth]{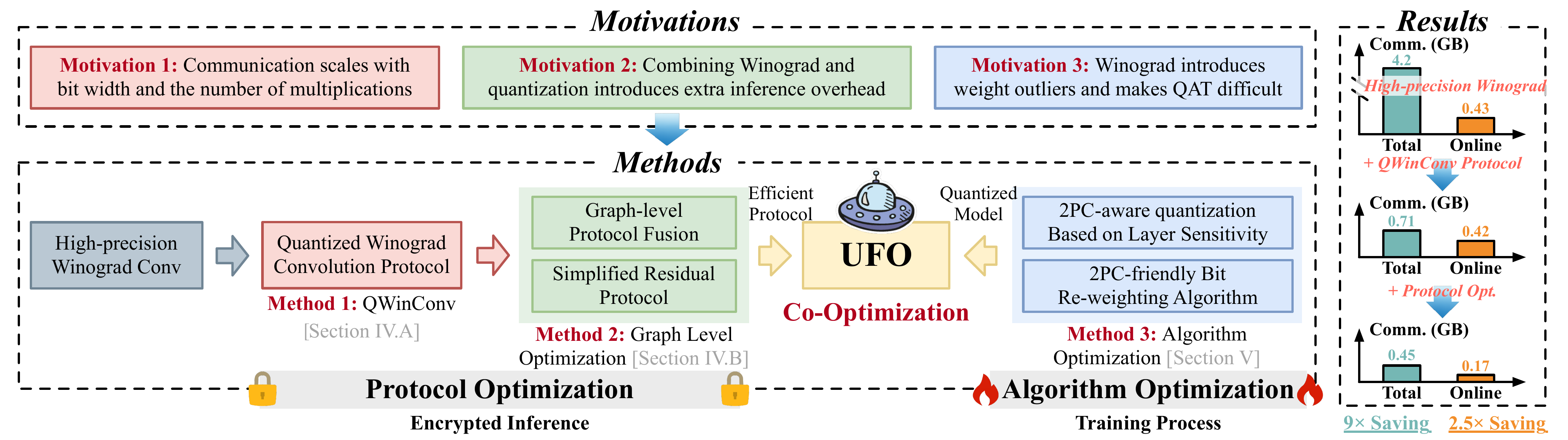}
    \caption{
    % \wx{Contribution should be more concise.}
    Motivations and our methods of \method.
    % Overview of \method~and the communication cost after each optimization step.
    Example result is evaluated on ResNet-32 with W2A6, eventually achieving 9$\times$ and 2.5$\times$ total and online communication reduction. 
    % \wx{The figure is too complex}
    }
    \label{fig:overview}
\end{figure*}

\begin{table*}[!tb]
\centering
\caption{Comparison with existing works. ``\checkmark'' means this term can be optimized. \# MUL means the number of multiplications, CPM means communication per MUL, and MP means mixed precision.}
\label{tab:compare_exist}
\resizebox{\linewidth}{!}{
\begin{tabular}{c|cc|c|cc|c}
\toprule
\multirow{2}{*}{\tabincell{c}{\textbf{Representative} \textbf{Method}}} & \multicolumn{2}{c|}{\textbf{Protocol Optimization}} & \multirow{2}{*}{\tabincell{c}{\textbf{Algorithm} \textbf{Optimization}}} & \multicolumn{2}{c|}{\textbf{Linear Optimization}} & \multirow{2}{*}{\tabincell{c}{\textbf{Non-linear} \textbf{Optimization}}}  \\
\cmidrule{2-3} \cmidrule{5-6}
                           & \textbf{Operator Level}    & \textbf{Graph Level}   &                               & \textbf{\# MUL $\downarrow$}  & \textbf{CPM $\downarrow$}  &                                   \\
\hline                           
SNL \cite{cho2022selective}, SENet \cite{kundu2023learning}, DeepReduce \cite{jha2021deepreduce}  & -  & -   & ReLU Pruning & -  & -   &   \checkmark   \\
\hline
SiRNN \cite{rathee2021sirnn}  &  Uniform Quant  &  MSB Opt  &  \tabincell{c}{-}   &  -    & \checkmark   &  \checkmark  \\
\hline
XONN \cite{riazi2019xonn}  &  Binary Quant     &   -    &  \tabincell{c}{\tabincell{c}{Binary Quant \\ (Large Accuracy Drop)}}    &   -  &   \checkmark    &   \checkmark  \\
% \hline
% \cite{hussain2021coinn}   & Factorized GEMM  & Protocol Conversion  &  \tabincell{c}{MP Quant}   & \checkmark  &  \checkmark   &  \checkmark   \\
\hline
ABNN2 \cite{shen2022abnn2} &  -  & -   & \tabincell{c}{Uniform Quant}  & -  &  \checkmark  &   \checkmark    \\
\hline
CoPriv \cite{zeng2023copriv}  & Winograd Conv  & - &  Re-param  &   \checkmark  & -  &  \checkmark \\
\hline
Quotient\cite{agrawal2019quotient}  &  -  &  -  &  \tabincell{c}{Gradient Quant}  &   -   &  \checkmark   &  \checkmark  \\
\hline
% \rowcolor[rgb]{ .949,  .953,  .961}
\rowcolor{orange!10}
\method~(ours)   &  \tabincell{c}{MP Quant \\ Winograd Conv}   &   \tabincell{c}{Simplified Residual, \\ Protocol Fusion, MSB Opt}    &  \tabincell{c}{2PC-aware Quant, \\ MP Quant, Bit Re-weighting}   &   \checkmark   &  \checkmark   &   \checkmark   \\
\bottomrule                       
\end{tabular}
}
\end{table*}

\subsubsection{\textbf{Motivation 2: a naive combination of Winograd convolution and quantization provides limited communication reduction for 2PC inference (induced by feature and output transformation)}}
Although Winograd convolution reduces the number of multiplications, it is not friendly to quantized inference
since it introduces many more local additions during the feature and output transformation.
Hence, to guarantee the computation correctness with avoiding overflow, extra bit width conversion protocols are required
as shown in Figure~\ref{fig:compare_extend}. Take ResNet-32 with 2-bit weight and 6-bit activation (W2A6) as an example in Figure \ref{fig:winograd_breakdown},
naively combining Winograd convolution and quantization achieves only $\sim$20\% communication reduction compared to not using Winograd convolution in SiRNN \cite{rathee2021sirnn},
which is far less compared to the benefit of Winograd convolution \cite{zeng2023copriv}.
Therefore, further protocol optimization is crucial to reduce the overhead induced by the bit width conversions.

\subsubsection{\textbf{Motivation 3: Winograd transformations introduce weight outliers and make low-precision QAT difficult (induced by weight transformation)}}
Since the weight transformation involves multiplying or dividing large plaintext integers \cite{lavin2016fast} and tends to
generate large weight outliers, it makes the quantization challenging.
Specifically, large outliers dominate the maximum magnitude measurement, leading to large quantization errors and inferior model accuracy.
We show the weight distributions and z-scores with and without Winograd transformations in Figure~\ref{fig:intro_bar}(c) and (d).
Instead of simply increasing the quantization bit width like previous works, we observe that it is possible to accommodate
the weight outliers without explicitly increasing the bit width.
% Recall for OT-based linear layers, 
Since each weight is written as $\sum_{b=0}^{l_w-1} w^{(b)} \cdot 2^b$ in OT-based linear layers,
% \footnote{we ignore the sign bit for convenience}
% and then, each bit $w^{(b)}$ is multiplied with the corresponding activations with a single OT. 
% The final results are acquired by combining the partial results by shift and addition.
% This provides us with unique opportunities to 
we can re-weight each bit by adjusting $2^b$ to increase the representation range without causing extra communication overhead.
In Figure~\ref{fig:skip}, bit re-weighting increases the representation range with the same bit width (note the total number of possible represented values remains the same).
% Bit re-weighting helps accommodate the weight outliers and improves the accuracy during QAT.

\subsubsection{\textbf{Paper Organization}}
\label{subsec:overview}
Based on the motivations, we propose \method, a protocol-algorithm co-optimization framework for efficient 2PC inference.
We show the overview in Figure \ref{fig:overview}.
% First, we optimize the 2PC protocols at both operator and graph level. 
% We first optimize the 2PC protocol for convolutions combining quantization and Winograd transformation (Section \ref{sec:winograd}).
We first build a novel convolution protocol that combines quantization and Winograd convolution (Section \ref{sec:winograd}).
We then propose a series of graph-level optimizations, including protocol fusion to reduce the communication for Winograd transformations (Section \ref{sec:fusion}),
simplified residual protocol to remove redundant bit width conversion protocols
(Section \ref{sec:fusion}), and 
% activation sign propagation and
protocol optimization given known most significant bits (MSBs)
(Section \ref{sec:msb}).
% Even Winograd is quantized to low precision, there is no benefit for online communication due to the extra bit width conversions.
% Our graph-level optimizations enable to reduce
% the online communication by 2.5$\times$ in the example.
We propose a 2PC-aware mixed-precision QAT algorithm and bit re-weighting algorithm to enhance the model accuracy (Section \ref{sec:quant}).
% network optimization to enable accurate yet efficient private inference via a communication-aware bit width assignment algorithm 
% \method~eventually reduces the total communication by 9$\times$ in the example.
We show a case evaluated on ResNet-32 with W2A6, eventually achieving 9$\times$ and 2.5$\times$ total and online communication reduction.

\section{Inference Protocol Optimization}
\label{sec:method}

% For the residual addition, we simplify the protocol that is suitable for \method~(Section \ref{sec:residual}).
% At the graph level, we introduce MSB-known optimization to reduce the communication of re-quantization and extension when the most significant bits (MSBs) of the intermediate values are known (Section \ref{sec:msb}).
% We further propose network optimization to enable accurate yet efficient quantized inference via a communication-aware bit width assignment algorithm (Section \ref{sec:qat}).
% With the protocol-network co-optimization, \method~reduces the convolution communication by 14.5$\times$ in the example.
% \ml{Have we introduced online and pre-processing stages before?\checkmark Section 2.3}

\begin{figure*}[!tb]
    \centering
    \includegraphics[width=\linewidth]{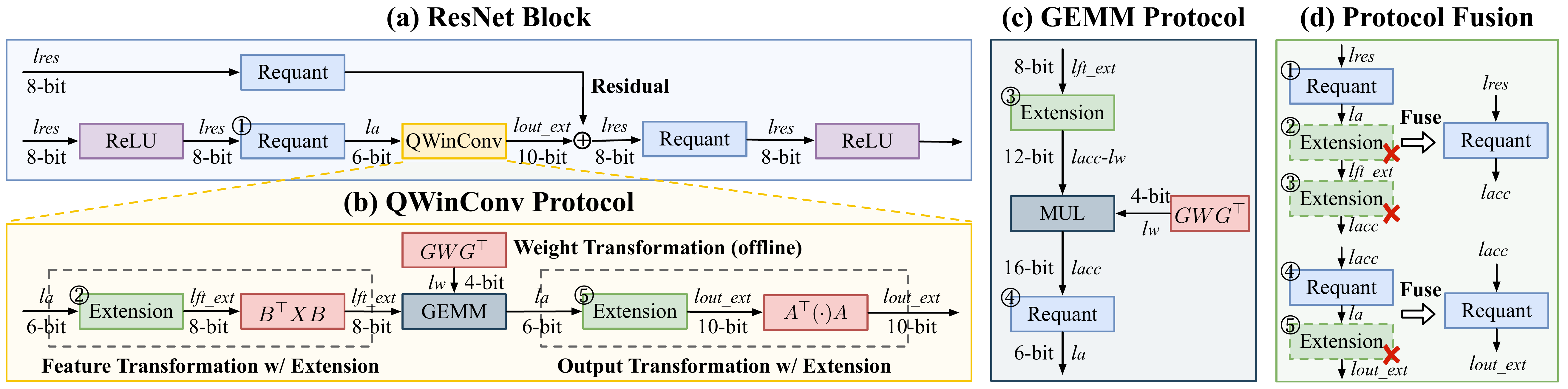}
    \caption{Overall protocol framework with our proposed quantized Winograd convolution. $l$ denotes the bit width and we give an example of W4A6.
    % ~in (a) an example residual block; (b) the design of QWinConv; (c) GEMM protocol in SiRNN \cite{rathee2021sirnn}; (d) graph-level protocol fusion.
    % \wx{How to simplify?}
    }
    \label{fig:protocol}
    % \vspace{-15pt}
\end{figure*}

\begin{figure}[!tb]
    \centering
    \includegraphics[width=\linewidth]{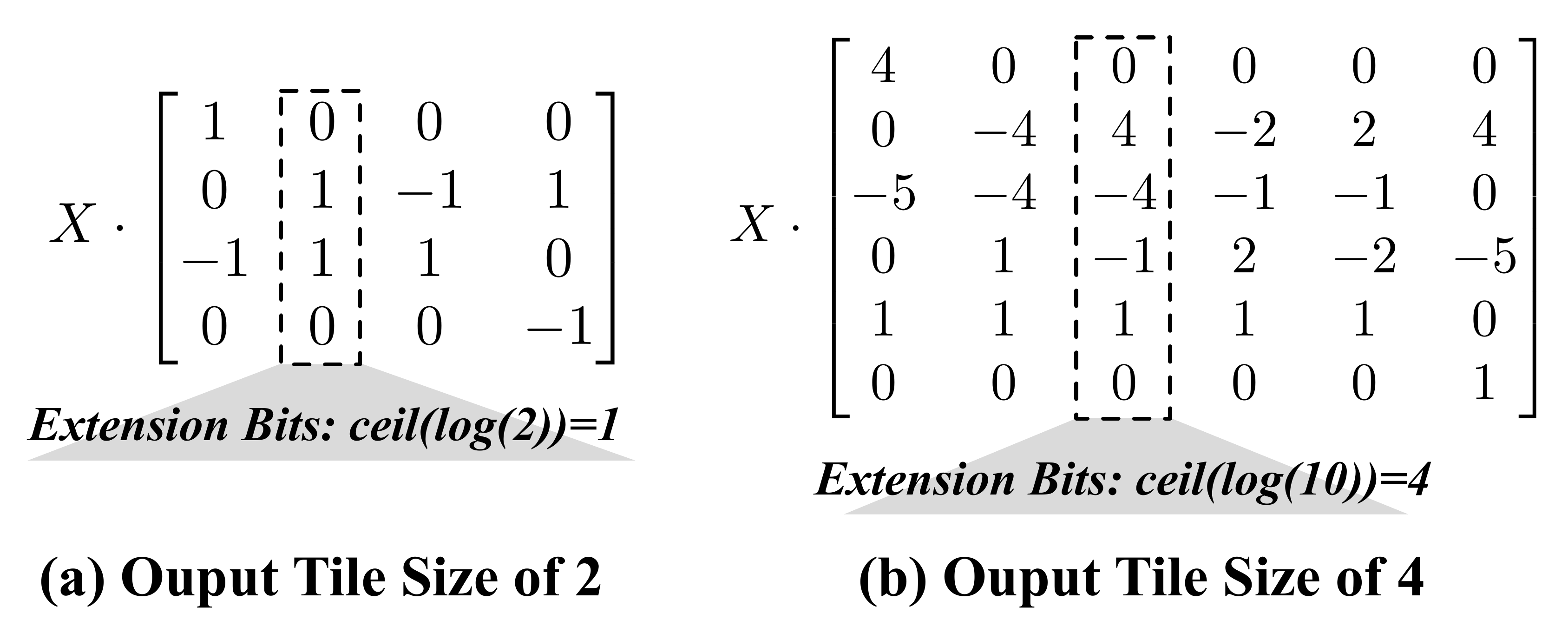}
    \caption{$XB$ in feature transformation. Dash boxes are the possible columns with the maximum number of additions. As a result, 2-bit and 8-bit extensions are needed for $B^\top XB$.}
    \label{fig:matmul_add}
    % \vspace{-15pt}
\end{figure}

\subsection{New Protocol: Quantized Winograd Convolution}
\label{sec:winograd}

% \zwx{Unify MatMul and GEMM}
% \ml{When writing the techniques, need to highlight: 1) how this is related to our observation/motivation? 2)
% what is the major non-trivial point of the technique; 3) describe the techniques; and 4) make quantitative comparison.}

% The logic of writing techniques (challenge-driven):
% 1) Previous works consider the local additions have no impact on private inference.
% 2) But it is not right since the bit width change.
% 3) So, we need extra bit extension.
% 4) Non-trivial points: a) larger $m$, larger bit width (comparison and can have a Fig. of matmul); b) where do we insert extension.
% 5) We choose to use $m=2$ and the insertion way.

% \ml{The symbols of transformation matrix $B$ and the bit width $B_w$ have conflicts.\checkmark\zwx{Change to l_w and l_x.}}

As explained in Section~\ref{sec:motivation}, when combining the Winograd convolution and quantization,
extra bit extensions are needed to strictly guarantee output correctness due to local additions in the feature and
output transformations.

\subsubsection{\textbf{Protocol Framework}}
% Based on Motivation 1 in Section \ref{sec:motivation}, 
We carefully design a 2PC framework with the quantized Winograd convolution protocol named QWinConv in Figure \ref{fig:protocol}.
In Figure \ref{fig:protocol}(a), before QWinConv, following SiRNN \cite{rathee2021sirnn},
activations are first re-quantized to $l_a$-bit to reduce the communication of QWinConv (\textcolor{black}{\ding{172}}).
% \ml{WinConv not defined\checkmark},
After QWinConv, the residual and QWinConv output are added together with the residual protocol.
Residuals are always kept in high bit width, i.e., 8-bit, for better accuracy \cite{yang2021fracbits}.
The detailed design of QWinConv is shown in Figure \ref{fig:protocol}(b), which involves 4 steps:
feature transformation, weight transformation, Winograd-domain GEMM, and output transformation.
We insert the bit extension protocols right before the feature and output transformation to avoid inference overflow as marked with \textcolor{black}{\ding{173}} and \textcolor{black}{\ding{176}}.
The GEMM protocol in Figure \ref{fig:protocol}(c) follows SiRNN \cite{rathee2021sirnn}
and the bit extension (\textcolor{black}{\ding{174}}) ensures accumulation correctness.
Note that weight transformation and its quantization are conducted offline,
while feature transformation and output transformation are executed online.
% We insert bit extensions right before the feature and output transformation to avoid inference overflow as marked with \textcolor{black}{\ding{183}} and \textcolor{black}{\ding{186}}, respectively.
% Extra bit extensions for Winograd transformation are marked as block \ding{173} and \ding{175}.
% As mentioned in Motivation 2 in Section \ref{sec:motivation}, although the Winograd transformations can be conducted locally, bit width extensions are needed to prevent overflow.
% although Winograd transformation leads to more local additions, these additions can impact the bit width of the operands in the quantized private inference.
% Hence, to ensure the inference correctness, we reserve enough bit width for the Winograd transformation in advance.
% To this end, extensions before transformation are required, which are marked as \ding{173} and \ding{174}.
% Considering the extreme condition, 2-bit and 4-bit extensions are required to guarantee the correctness of transformation for $F(2\times 2, 3\times 3)$.
% Hence, $B^TXB$ is computed with 8-bit precision and $A^T(\cdot)A$ is computed with 10-bit precision, respectively.
% \zwx{Mention that 1) the communication scales with tile size $r$, and $r=2$ is friendly to \method; 2) different insertion of extension.}

\subsubsection{\textbf{Approximation of Extension Bits and Tile Size Selection}} 
A natural question is \textit{``how many bits to extend for Winograd transformations while ensuring the inference correctness?''}
Since transformation matrices are constant and related to the tile size,
we have the following lemma that bounds the outputs after transformation.
% Two natural questions hence arise: 
% \textit{
% \underline{1)} where to insert the bit width extension protocols? \underline{2)} how many bits to extend?
% }

% \wx{Weaken the first discussion.}
% For the first question, there are different ways to insert the bit extension. 
% Take the output transformation as an example. 
% One method is to extend the activations right before computing the output transformation,
% which incurs online communication.
% The second method is to insert the bit width extensions before the GEMM protocol.
% While this enables to merge the bit extension protocols with the offline weight transformation,
% we find it will drastically increase the GEMM communication and thus, choose the first method in \method.

% For the second question, we have the following lemma that bounds the outputs after transformation.

\begin{lemma}
    Consider $Y = XB$ in feature transformation, where $B$ is the transformation matrix. For each element $Y_{i, j}$, its magnitude can always be bounded by
    % For an inner product of $i$-th row of $X$ and $j$-th column of transformation matrix $B$, denoted as $X_{i, :}$ and $B_{:, j}$.
    \begin{align*}
    |Y_{i, j}| = |\sum_{k} X_{i, k} B_{k, j}| & \leq \sum_{k} | X_{i, k} | |B_{k, j}|  \\
                                  &\leq 2^{l_x} \sum_k |B_{k, j}| \leq 2^{l_x + \lceil\log_2 ||B_{:, j}||_1\rceil},
    \end{align*}
    where $||\cdot||_1$ is the $\ell_1$-norm, $\lceil\cdot\rceil$ is the ceiling operation.
\end{lemma}

Therefore, each output element $Y_{i, j}$ requires at most $l_x + \lceil\log_2 ||B_{:, j}||_1\rceil$ bits.
Since we use per-tensor activation quantization, to strictly guarantee output correctness, we need
$\lceil\max_{j} l_x + \log_2 ||B_{:, j}||_1\rceil$ bits to represent the output. 
% Similarly, to compute $B^\top X$, $\max_{j} \log_2 ||B_{:, j}||_1$ is needed, adding up to $2 \times \max_{j} \log_2 ||B_{:, j}||_1$ bits to extend. 
Hence, $2 \cdot \lceil\max_{j} \log_2 ||B_{:, j}||_1\rceil$ bits are required to extend for $B^\top XB$.
In Figure \ref{fig:matmul_add}, we show the transformation matrix $B$ with the output tile size of 2 and 4. As can be calculated,
2-bit and 8-bit extensions are needed for $B^\top XB$, respectively. 
As 8-bit extension drastically increases the communication
of the following GEMM, we choose 2 as the output tile size. 
Output transformation can be calculated similarly and 4-bit extension is needed for $A^\top [\cdot] A$ with the output tile size of 2.

% Based on Motivation 1 in Section \ref{sec:motivation}, 
Based on the above analysis, we propose the quantized Winograd convolution protocol, dubbed QWinConv, in Figure \ref{fig:protocol}.
% The framework consists of three parts: (a) an example residual block; (b) our proposed quantized Winograd-based convolution protocol named WinConv; and (c) MatMul protocol.
% Take W4A6 (4-bit weight and 6-bit activation) as an example. 
Consider an example of 4-bit weights and 6-bit activations.
In Figure \ref{fig:protocol}(a), before QWinConv, following SiRNN \cite{rathee2021sirnn},
activations are first re-quantized to 6-bit to reduce the communication of QWinConv (block \ding{172}).
% \ml{WinConv not defined\checkmark},
We always keep the residual in higher bit width, i.e., 8-bit, for better accuracy \cite{wu2018mixed,yang2021fracbits}.
After QWinConv, the residual and QWinConv output are added together with the residual protocol.
% \ml{The EWMM to GEMM conversion is already discussed previously.\checkmark No need to have Wrap \& MUX separately. Maybe combine CrossTerm and Wrap to a new name. Need to point out GEMM follows SiRNN and the extension is for the accumulation correctness.\checkmark The description of the protocol needs to be improved.}
The design of QWinConv is shown in Figure \ref{fig:protocol}(b), which involves four steps:
feature transformation, weight transformation, Winograd-domain GEMM, and output transformation.
% Following CoPriv \cite{zeng2023copriv}, EWMM is equivalently converted to GEMM with tile aggregation to leverage the batch optimization, and GEMM is equipped with DNN architecture-aware adaptive protocol optimization based on the layer dimension.
The GEMM protocol in Figure \ref{fig:protocol}(c) follows SiRNN \cite{rathee2021sirnn}
and the extension (block \ding{174}) ensures accumulation correctness.
% Also, following \cite{zeng2023copriv}, GEMM is equipped with adaptive convolution optimization which selects the optimal OT sender based on the layer dimension. \ml{Not necessary.}
% It is worth noting that 
Weight transformation and its quantization can be conducted offline,
while feature and output transformation and quantization must be executed online.
We insert bit extensions right before the feature and output transformation (block \ding{173} and \ding{175}).

\subsection{Graph-level Protocol Optimization}

% \begin{figure}
%     \centering
%     \includegraphics[width=0.7\linewidth]{figure/bit_representation.pdf}
%     \caption{Conversion of bit representation with bit width $l_w=4$. During training, each bit in $W_s^{(3:0)}$ is trainable.}
%     \label{fig:bit_repre}
% \end{figure}

\subsubsection{\textbf{Graph-level Protocol Fusion}}
\label{sec:fusion}

As explained in Section~\ref{sec:motivation}, extra bit width conversion protocols in QWinConv increase the online
communication and diminish the benefits of combining quantization with Winograd convolution,
especially for low bit widths, e.g., 4-bit.
We observe there are neighboring bit width conversion protocols that can be fused to reduce the overhead in Figure \ref{fig:protocol} (more details in Section \ref{supp:fusion}).
Specifically, we find two patterns that appear frequently:
\underline{1)} consecutive bit width conversions, e.g., \textcolor{black}{\ding{172}}\textcolor{black}{\ding{173}}, \textcolor{black}{\ding{175}}\textcolor{black}{\ding{176}};
\underline{2)} bit width conversions that are only separated by local operations, e.g., \textcolor{black}{\ding{173}}\textcolor{black}{\ding{174}}.
As the cost of bit width conversion protocols is only determined by the bit width of inputs \cite{rathee2021sirnn} as shown in Table \ref{tab:protocols},
such protocol fusion enables complete removal of the cost of \textcolor{black}{\ding{173}}\textcolor{black}{\ding{174}}\textcolor{black}{\ding{176}} as shown in Figure \ref{fig:protocol}(d),
enabling efficient combination of Winograd convolution and quantization.

\begin{figure}[!tb]
    \centering
    \includegraphics[width=\linewidth]{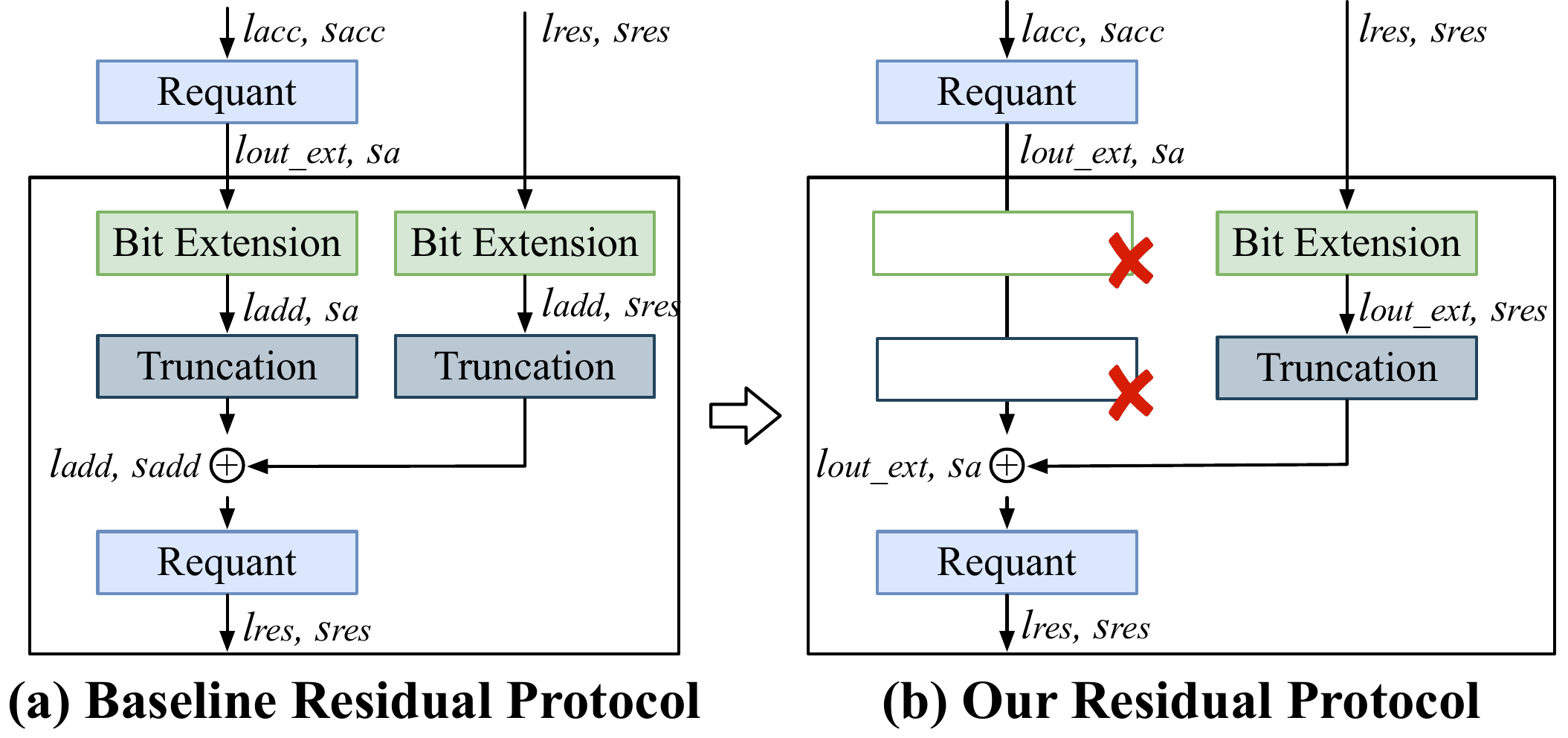}
    \caption{Comparison between (a) baseline residual protocol \cite{rathee2021sirnn} and (b) our simplified residual protocol. $l, s$ mean the bit width and scale of the operand. Our protocol removes redundant bit width conversion in the main branch.}
    \label{fig:residual_protocol}
    % \vspace{-15pt}
\end{figure}

\subsubsection{\textbf{Simplified Residual Protocol}}
\label{sec:residual}
% \ml{For all these techniques, it is unclear why they are needed. More motivations and complexity comparisons are needed.}
% \ml{Do we want to introduce quantization in the preliminary?\checkmark}
As shown in Figure \ref{fig:intro_bar}(b), residual protocol consumes around 50\% of the online communication due to the complex alignment 
of both bit widths and scales
% \footnote{We explain the details of bit width and scales in quantization in Appendix \ref{supp:quant}.} 
as shown in Figure \ref{fig:residual_protocol}(a).
Therefore, we propose a new simplified residual protocol to reduce the communication as shown in Figure \ref{fig:residual_protocol}(b).
% and meanwhile, keep the advantage of protocol fusion.
% \begin{figure}[!tb]
% \begin{wrapfigure}{r}{0.45\textwidth}
%     % \vspace{-20pt}
%     \centering
%     \includegraphics[width=\linewidth]{figure/residual.pdf}
%     \caption{Comparison between (a) baseline residual protocol \cite{rathee2021sirnn} and (b) our simplified residual protocol. $l, s$ mean the bit width and scale of the operand, respectively.}
%     \label{fig:res_protocol}
%     % \vspace{-20pt}
% \end{wrapfigure}
Specifically, we directly align the bit width and scale of residual to the QWinConv output for addition.
In this way, we get rid of the redundant bit width conversion protocols for the QWinConv output.
As demonstrated in Section \ref{sec:micro}, \method~drastically reduces communication costs.

% reducing the communication from $O(\lambda(l_{out\_ext}+l_{res}+2l_{add}+8))$ to $O(\lambda(l_{out\_ext}+l_{res}+4))$.
% but also keep the opportunity of fusing the neighboring re-quantization and extension of block \ding{175} and \ding{176} in Figure \ref{fig:protocol}.
% \ml{\ding{175} and \ding{176} are already merged?\checkmark}
% \ml{Which one?}

% We also note there is another way to simplify the residual protocol.
% We can skip the top re-quantization, however, this way leads to a much larger bit width and cannot support for the fusion of re-quantizaiton and extension.

\subsubsection{\textbf{MSB-known Optimization}}
\label{sec:msb}
As pointed out by \cite{huang2022cheetah,hao2022iron}, 2PC protocols can be more efficient when the MSB of the secret shares is known.
Since the activation after ReLU is always non-negative, protocols including truncation and extension can be further optimized.
% previous works including Cheetah \cite{huang2022cheetah} proposes most significant bit (MSB) optimization for communication reduction.
In \method, we locate the ReLU function and then, propagate the sign of the activations to all the downstream protocols.
In Figure \ref{fig:protocol}, for example, the input of \textcolor{black}{\ding{172}\ding{173}\ding{174}} must be non-negative,
so they can be optimized. 
In contrast, \textcolor{black}{\ding{175}\ding{176}} can not be optimized since the GEMM outputs can be either positive or negative.
% With known MSB, the communication of $\Pi_{\mathrm{Ext}}^{l_1, l_2}$ is reduced from $O(\lambda(l_1+1)+13l_1+l_2)$ to $O(2\lambda-l_1+l_2+2)$ and $\Pi_{\mathrm{Trunc}}^{l_1, l_2}$ is reduced from $O(\lambda(l_1+3)+15l_1+l_2+20)$ to $O(3\lambda+l_1+l_2+20)$.
% \zwx{Needs to be simplied}

\subsection{Complexity and Formal Description of Protocol Fusion}
\label{supp:fusion}

\subsubsection{\textbf{Complexity Analysis}}
The communication complexity per element mainly scales with the initial bit width $l_1$.
Also, for a given matrix with the dimension of $d_1\times d_1\times d_3$, the number of extensions needed would be $d_1\times d_1\times d_3$.
Therefore, the total communication complexity of extension for a matrix becomes $O(d_1d_2d_3(\lambda (l_1+1)+13l_1+l_2))$.
In Figure \ref{fig:protocol}, the extension communication of feature extension (block \ding{173}) is $O(HWC(\lambda (l_1+1)+13l_1+l_2))$ while that of output extension (block \ding{176}) is $O(KT(m+r-1)^2(\lambda (l_1+1)+13l_1+l_2))$, which is more expensive.

\subsubsection{\textbf{Protocol Fusion}}
We provide the following propositions to elaborate the principle of protocol fusion.
\begin{proposition}
\label{prop:decomp}
    % \ml{1) This proposition is hard to understand; 2) the way the proposition is described is not correct; 3) need to analyze the communication complexity before and after the decomposition; 4) re-quant = Trunc?}
    % Truncation is widely used in quantized inference to avoid numerical overflow \cite{rathee2020cryptflow2,rathee2021sirnn,zeng2023copriv}.
    For a given $\langle x\rangle^{(l_1)}, 
    \Pi_{\mathrm{Trunc}}^{l_1, l_2}\left(\langle x\rangle^{(l_1)}\right)$ can be \\[0.5em] decomposed into $\Pi_{\mathrm{TR}}^{l_1, l_2}$ followed by $\Pi_{\mathrm{Ext}}^{l_1-l_2, l_1}$ as
    \begin{align*}
        \Pi_{\mathrm{Trunc}}^{l_1, l_2}\left(\langle x\rangle^{(l_1)}\right) & =  \Pi_{\mathrm{Ext}}^{l_1 - l_2, l_1}\left( \Pi_{\mathrm{TR}}^{l_1, l_2}\left(\langle x\rangle^{(l_1)}\right)\right).
    \end{align*}
    The decomposition does not change communication.
    % Decomposition reduces the communication from $O(\lambda(l_1+3))$ to $O(\lambda(l_1+2))$.
\end{proposition}

\begin{proposition}
\label{prop:ext}
    If a given $\langle x\rangle^{(l_1)}$ is extended to $l_2$-bit first and then extended to $l_3$-bit. Then, the two neighboring extensions can be fused together as
    \begin{align*}
        \Pi_{\mathrm{Ext}}^{l_2, l_3}\left(\Pi_{\mathrm{Ext}}^{l_1, l_2}\left(\langle x\rangle^{(l_1)}\right)\right) & = \Pi_{\mathrm{Ext}}^{l_1, l_3}\left(\langle x\rangle^{(l_1)}\right).
    \end{align*}
    Extension fusion reduces communication from $O(\lambda(l_1+l_2+2))$ to $O(\lambda(l_1+1))$. 
    % \ml{Is this correct? independent of $l_2$?\checkmark}
\end{proposition}

\begin{proposition}
    % \ml{Need more explanations here. TR does not show up in previous propositions.}
    % Combining two propositions, 
    % If a given $\langle x\rangle^{(l_1)}$ is first truncated to $l_2$-bit and then extended to $l_3$-bit. Then, the protocol can be
    % re-combined as 
    For a given $\langle x\rangle^{(l_1)}$,
    when re-quantization ends up with truncation, and is followed by an extension, the protocol can be first decomposed and then fused as
    % if an extension follows re-quantization, truncation can be computed as
    \begin{align*}
        & \Pi_{\mathrm{Ext}}^{l_1, l_3}  \left(\Pi_{\mathrm{Trunc}}^{l_1, l_2}\left(\langle x\rangle^{(l_1)}\right) \right) 
        \\[0.5em] =&
        \Pi_{\mathrm{Ext}}^{l_1, l_3} \left( \Pi_{\mathrm{Ext}}^{l_1 - l_2, l_1}\left( \Pi_{\mathrm{TR}}^{l_1, l_2}\left(\langle x\rangle^{(l_1)}\right)\right)\right) 
        \\[0.5em] =& \Pi_{\mathrm{Ext}}^{l_1 - l_2, l_3}\left( \Pi_{\mathrm{TR}}^{l_1, l_2}\left(\langle x\rangle^{(l_1)}\right)\right).
    \end{align*}
    Combining Proposition \ref{prop:decomp} and \ref{prop:ext}, this fusion reduces communication by around 2$\times$ from $O(\lambda(2l_1+4))$ to $O(\lambda(l_1+2))$.
\end{proposition}

\subsection{Security and Correctness Analysis of \method}
\label{sec:security}
\method~is built on top of SiRNN \cite{rathee2021sirnn} with a new quantized Winograd convolution protocol and graph-level protocol optimizations. The security of the QWinConv protocol directly follows
the security guarantee from OT primitive \cite{naor2001efficient}. Observe that all the communication between the client and the server is performed through OT which ensures the privacy of both the selection bits and messages. Graph-level protocol optimizations
% ,including protocol fusion, residual simplification, and MSB-known optimization 
leverage information known to both parties,
e.g., the model architecture, and thus, do not reveal extra information.
Utilizing quantized neural networks does not affect OT primitives in any way.
The correctness of the QWinConv protocol is guaranteed by the theorem of Winograd transformation
\cite{lavin2016fast} and we use bit extensions to avoid inference errors.

% adversaries $\mathcal{A}$. 
% Formally, our security analysis follows the widely accepted simulation paradigm targeting static semi-honest probabilistic polynomial time (PPT) adversaries. The core principle is to prove that the view of an adversary $\mathcal{A}$ in a real execution is indistinguishable from a simulated view in an ideal execution, where the computation is handled by a trusted functionality. Specifically, security holds if a simulator $\mathcal{S}$ can be constructed in the ideal setting such that no environment $\mathcal{Z}$ can differentiate between the real and ideal interactions. Furthermore, we employ the hybrid model for modularity, treating sub-protocols as calls to ideal functionalities F within our analysis.

\section{Quantization Algorithm Optimization}
\label{sec:quant}

In this section, we propose a Winograd-domain QAT algorithm that is compatible with OT-based 2PC inference.
The training procedure is shown in Algorithm \ref{alg:quant}.
We first assign layer-wise bit widths based on quantization sensitivity (line 2) and then propose 2PC-friendly bit re-weighting to consider weight outliers (line 3-8).
Then, training is performed (line 9-14).
We demonstrate where to insert the quantization operation in Figure \ref{fig:winograd_qat}.
Then, 2PC inference based on optimized protocols and re-weighted bit importance is performed in Algorithm \ref{alg:infer_details}.

\subsection{2PC-aware Sensitivity-based Bit Width Assignment}
\label{sec:qat}

% \noindent \textbf{Hessian-based Bit Width Assignment}
Different layers in a CNN have different sensitivity to the quantization noise \cite{dong2019hawq,dong2020hawq,yao2021hawq}.
To enable accurate and efficient private inference of \method, 
we propose a 2PC-aware mixed-precision algorithm to search for the optimal bit width for each layer based on sensitivity.
Following HAWQ \cite{dong2019hawq,dong2020hawq,yao2021hawq}, sensitivity can be approximately measured by the average trace of the Hessian matrix, and layers with a larger trace of the Hessian matrix are more sensitive to quantization.
Let $\Omega_i$ denote the output perturbation induced by quantizing the $i$-th layer. Then, we have
% The key insight is that layers with a larger trace of the Hessian matrix are more sensitive to quantization. 
% the perturbation of the $i$-th layer, denoted as $\Omega_i$, from the quantization error can be computed as
% \begin{align*}
$$
 \Omega_i = \overline{Tr}(H_i)\cdot ||\mathrm{Q}(GW_iG^\top)-GW_iG^\top||_2^2,
$$
% \end{align*}
where $H_i$ and $\overline{Tr}(H_i)$ denote the Hessian matrix and average trace,
$||\mathrm{Q}(GW_iG^\top)-GW_iG^\top||_2^2$ denotes the $L_2$ norm of Winograd quantization perturbation. 
% $\mathrm{Quant}(\cdot)$ denotes quantization function,
% and $||\cdot||_2$ denotes the $\ell_2$-norm of quantization perturbation.
Now we consider the factor of 2PC communication cost.
Given the communication bound $\zeta$ and an $L$-layer model,
we formulate the 2PC-aware bit width assignment problem as an integer linear programming problem (ILP) as
\begin{align*}
    \min_{\{b_i\}_{i=1}^L} \sum_{i=1}^L \Omega_{i}^{b_i},
    ~~\mathrm{s.t.} \sum_{i=1}^L C_{i}^{b_i} \leq \zeta,
\end{align*}
where $C_{i}^{b_i}$ is the associated communication cost and $\Omega_{i}^{b_i}$ is perturbation when
quantizing the $i$-th layer to $b_i$-bit.
The objective is to find the optimal bit width assignment $\{b_i\}_{i=1}^L$ to minimize the perturbation of the model output under the given communication constraint.

\begin{algorithm}[!tb]
\caption{Training Procedure of \method~Quantization with Mixed Precision}
\label{alg:quant}
\SetKwInOut{Input}{Input}
\SetKwInOut{Output}{Output}
\SetKwFunction{Tuple}{}
\SetKwComment{Comment}{/* }{ */}

\Input{Pre-trained floating-point weights $w_f$, communication bound $\zeta$, the number of layers $L$, finetuning epochs $E$.}
\Output{Set of bit importance $\mathbb{B}$ and quantized weights $w$.}
\BlankLine

\tcp{Layer-wise bit width assignment (Sec. \ref{sec:qat})}
$l_w = \mathrm{HessianAlgoritm}(w_f, \zeta)$; 
% \hfill\textcolor{black}{$\triangleright$ Layer-wise bit width assignment}
% \For{$l \in [0, \ldots, L - 1]$}
% {
% }

\For{$i \in [1, \ldots, L]$}
{
    % $O^{(l)} = \texttt{has\_{outlier}}(w)$; // Determine whether the layer has large outliers
    \tcp{Original bit importance}
    $\mathbb{B}_{i} \leftarrow \{2^{l_{wi-1}}, 2^{l_{wi-2}} \ldots, 2^1, 2^0\}$; 
    % \hfill\textcolor{black}{$\triangleright$ Original bit importance}
    
    \If{$\mathrm{HasOutliers}(w_{i})$}  
    {
        \tcp{Bit re-weighting (Sec. \ref{sec:bit_reweighting})}
        $\mathbb{B}_{i} \leftarrow \{2^{l_{wi}}, 2^{l_{wi-2}}, \ldots, 2^1, 2^0\}$; 
        % \hfill\textcolor{black}{$\triangleright$ Bit re-weighting}
    }
}

\For{$e \in [1, \ldots, E]$} 
{
    \tcp{Training (Sec. \ref{sec:finetuning})}
    % \textbf{Forward:} $w_q=s\cdot(\sum_{b=0}^{l_w-1}w^{(b)}\cdot \mathbb{B}[b])/(2^{l_w}-1)$ \Comment*[r]{Re-weighted quantization} 
    % Compute loss $\mathcal{L}$ with cross-entropy loss; \\
    % \textbf{Backward:} $\frac{\partial \mathcal{L}}{\partial w^{(b)}} = \frac{2^b}{2^{l_w}-1} \frac{\partial \mathcal{L}}{\partial w_q}$ with STE \cite{bengio2013estimating}; \\
    % Update $w^{(b)}$ with SGD optimizer;
    Forward propagation based on Equation (\ref{eq:bsq_forward}); 
    % \hfill\textcolor{teal}{$\triangleright$ Re-weighted computation}
    
    Compute loss $\mathcal{L}_{CE}$; \\
    
    Back propagation based on Equation (\ref{eq:bsq_back}); 
    % \hfill\textcolor{black}{$\triangleright$ STE}
    
    Update $w$ with SGD optimizer;
}

\KwRet $w$ and $\mathbb{B}$.

\end{algorithm}

\begin{algorithm*}[!t]
\caption{Detailed 2PC Inference Protocol of \method~for a Single ResNet Block}
\label{alg:infer_details}
% 保持您喜欢的简洁风格，只定义必要的 Block
\SetKwInOut{Input}{Input}
\SetKwInOut{Output}{Output}
\SetKwBlock{Offline}{\textcolor{red!60!black}{[Offline Phase]} Weight Transformation and Quantization}{end}
\SetKwBlock{Online}{\textcolor{red!60!black}{[Online Phase]} Winograd-based Convolution, Non-linear Layer, Residual}{end}
\SetKwComment{tcp}{// }{}

\Input{Secret-shared input $\langle \mathbf{X}\rangle \in \mathbb{Z}_{2^{l_{in}}}^{N \times C \times H \times W}$, quantized weights $\mathbf{W}$, Winograd transformation matrices $\mathbf{A}, \mathbf{B}, \mathbf{G}$.}
\Output{Secret-shared output $\langle \mathbf{Y}\rangle \in \mathbb{Z}_{2^{l_{out}}}^{N \times K \times H^\prime \times W^\prime}$.}
\BlankLine

\Offline{
    % \tcp{Weight Transformation and Quantization}
    $\mathbf{W}_{win} \leftarrow \mathbf{G} \mathbf{W} \mathbf{G}^\top$ \tcp*[r]{Weight transformation (Eq. (\ref{eq:winograd}))}
    $\mathbf{W}_{q} \leftarrow \mathrm{Quant}(\mathbf{W}_{win})$ \tcp*[r]{Weight quantization with bit re-weighting (Sec. \ref{sec:quant})}
    % Generate Beaver triples for OT-based GEMM\;
    % compatible with shapes $(K \times C)$ and $(C \times T)$\;
}

\Online{
    \tcp{====================== Quantized Winograd Convolution ======================}
    \tcp{1. Feature Partitioning}
    $\langle \tilde{\mathbf{X}} \rangle \leftarrow \mathrm{Partition}(\langle \mathbf{X} \rangle)$ 
    % \tcp*[r]{Reshape: $N \times H \times W \to T \times n \times n$}
    
    \tcp{2. Feature Transformation (Local Computation)}
    $\langle \tilde{\mathbf{X}} \rangle_{ext} \leftarrow \Pi_{\mathrm{Ext}}(\langle \tilde{\mathbf{X}} \rangle)$ \tcp*[r]{Extend bit width to prevent overflow}
    $\langle \mathbf{U} \rangle \leftarrow \mathbf{B}^\top \langle \tilde{\mathbf{X}} \rangle_{ext} \mathbf{B}$ \tcp*[r]{Computed locally on each tile (Eq. (\ref{eq:winograd}))}
    
    \tcp{3. Winograd-domain GEMM}
    % 展开说明 GEMM 是如何进行的
    % \zwx{I think there is no need to explain all details}
    % View $\langle \mathbf{U} \rangle$ as $n^2$ matrices of shape $(C \times T)$\;
    % \For{$p \leftarrow 1$ \KwTo $n^2$}{
        % $\langle \mathbf{M}_p \rangle \leftarrow \Pi_{\mathrm{GEMM}}(\mathbf{W}_{q}^{(l)}[p], \langle \mathbf{U} \rangle[p])$ 
        % \tcp*[r]{$(K \times C) \times (C \times T) \to (K \times T)$}
    % }

    $\langle \mathbf{M} \rangle \leftarrow \Pi_{\mathrm{GEMM}}(\mathbf{W}_{q}^{(l)}, \langle \mathbf{U} \rangle)$ \tcp*[r]{OT-based GEMM with communication (Eq. (\ref{eq:winograd}))}
    
    % $\langle \mathbf{M} \rangle \leftarrow \mathrm{Reshape}(\{\langle \mathbf{M}_p \rangle\})$ \tcp*[r]{Result shape: $T \times K \times n \times n$}
    
    \tcp{4. Output Transformation (Local Computation)}
    $\langle \mathbf{M} \rangle_{ext} \leftarrow \Pi_{\mathrm{Ext}}(\langle \mathbf{M} \rangle)$ \tcp*[r]{Extend bit width to prevent overflow}
    $\langle \mathbf{Y} \rangle \leftarrow \mathbf{A}^\top \langle \mathbf{M} \rangle_{ext} \mathbf{A}$ \tcp*[r]{Computed locally on each tile (Eq. (\ref{eq:winograd}))}
    % \tcp*[r]{Tile size reduces: $n \times n \to m \times m$}
    
    % \tcp{5. Scatter}
    % $\langle \mathbf{Y}_{mid} \rangle \leftarrow \mathrm{Fold}(\langle \mathbf{Y}_{win} \rangle)$ \tcp*[r]{Reshape: $T \times K \times m \times m \to N \times K \times H \times W$}
    
    \BlankLine
    \tcp{============================= Non-Linear Layer =============================}
    $\langle \mathbf{Y} \rangle \leftarrow \mathrm{ReLU}(\langle \mathbf{Y} \rangle)$ \tcp*[r]{MSB-known optimization (Sec. \ref{sec:msb})}
    $\langle \mathbf{Y} \rangle \leftarrow \Pi_{\mathrm{Trunc}}(\langle \mathbf{Y} \rangle)$ \tcp*[r]{Re-quantize to $l_{in}$}
    
    \BlankLine
    \tcp{====================== Quantized Winograd Convolution ======================}
    Repeat step \# 5–15 with $\langle \mathbf{Y} \rangle$ as the input to obtain the output $\langle \mathbf{Y}_{\text{out}} \rangle$\;
    
    \BlankLine
    \tcp{======================= Simplified Residual Protocol =======================}
    $\langle \mathbf{X} \rangle_{ext} \leftarrow \Pi_{\mathrm{Ext}}(\langle \mathbf{X} \rangle)$ \tcp*[r]{Align output bit width and scale}
    $\langle \mathbf{Y} \rangle \leftarrow \langle \mathbf{Y}_{out} \rangle + \langle \mathbf{X} \rangle_{ext}$ \tcp*[r]{Addition for residual (\ref{sec:residual})}
}

\KwRet $\langle \mathbf{Y} \rangle$\;

\end{algorithm*}

\subsection{2PC-friendly Bit Re-weighting Algorithm for Outliers}
\label{sec:winquant+}

\subsubsection{\textbf{Bit Re-weighting}}
\label{sec:bit_reweighting}
Based on the above algorithm, we assign the bit width of each layer.
However, as mentioned in Section \ref{sec:motivation}, weight outliers introduced by Winograd transformations make quantization challenging, leading to inferior model accuracy.
For OT-based linear layers, each weight is first written as $\sum_{b=0}^{l_w-1} w^{(b)} \cdot 2^b$ and then,
each bit $w^{(b)}$ is multiplied with the corresponding activations with a single OT.
This provides us with opportunities to re-weight each bit by adjusting $2^b$ to increase
the representation range without explicitly increasing the bit width or causing extra communication overhead.
We define $2^b$ as the bit importance for $b$-th bit and define the set of $2^b$ as
$\mathbb{B}=\{2^{l_w-1}, 2^{l_w-2} \ldots, 2^1, 2^0\}$.
Then, the weight can be re-written as $\sum_{b=0}^{l_w-1} w^{(b)} \cdot \mathbb{B}[b]$.
We first perform a one-time inference to check whether there are outliers in each layer based on the z-score.
% determine whether there is a weight outlier situation for each layer based 
% on the z-score.
% on metrics like Z-score \cite{shiffler1988maximum}\footnote{Given a set of data $x$, $z-score(x)=(x-\mu)/\sigma$, where $\mu, \sigma$ are the mean and standard deviation.}. 
Then, for the layers with outliers, we re-weight the bit importance by adjusting $2^b$ to $2^{b^\prime}$ where $b^\prime>b$
to increase the representation range flexibly. In practice, we increase MSB by adjusting
$\mathbb{B}$ to $\{2^{l_w}, 2^{l_w-2} \ldots, 2^1, 2^0\}$ to accommodate the outliers.

\subsubsection{\textbf{Finetuning}}
\label{sec:finetuning}
After the bit re-weighting, we use QAT to finetune the quantized models. Since the bit importance is adjusted, we find it convenient to
utilize the bit-level quantization strategy \cite{yang2021bsq} for finetuning,
which considers each bit $w^{(b)}$ as a separate trainable parameter.
% Bit-level quantization with bit representation is demonstrated in Figure \ref{fig:bit_repre}.
% We follow the bit-level quantization strategy \cite{yang2021bsq} to finetune our low-precision networks.
Specifically, the quantized weight after bit re-weighting can be formulated as
\begin{equation}
\label{eq:bsq_forward}
    \textbf{Forward:}~
    w_q = \frac{s\cdot \mathrm{Round} (\sum_{b=0}^{l_w-1} w^{(b)}\cdot \mathbb{B}[b] )}  {2^{l_w}-1},
\end{equation}
\begin{equation}
\label{eq:bsq_back}
    \textbf{Backward:}~
    \frac{\partial \mathcal{L}}{\partial w^{(b)}} = 
    % \frac{2^b}{2^{l_w}-1} \frac{\partial \mathcal{L}}{\partial w_q} = 
    \frac{2^b}{2^{l_w}-1} \frac{\partial \mathcal{L}_{CE}(\mathcal{M}_{w_q}(x), y)}{\partial w_q},
\end{equation}
where $(x, y)$ denotes input-label pair, $s$ denotes the scaling factor, $\mathcal{L}_{CE}$ denotes cross-entropy loss, $\mathcal{M}_{w_q}$ denotes the model with quantized weight $w_q$.
% and $w_s^{(b)}$ denotes 1-bit of $w_s$.
During finetuning, both $w^{(b)}$ and $s$ are trainable via STE \cite{bengio2013estimating}. 

\subsubsection{\textbf{Inference}}
After finetuning, 
we obtain each layer's quantized weights and the adjusted bit importance $\mathbb B$. 
The inference pipeline is shown in Algorithm \ref{alg:infer_details}.
For the $i$-th layer, QWinConv 
% takes secret shared activation, quantized weight $w_{i}$, and bit importance $\mathbb{B}_{i}$ as input.
represents weight as $w_i=\sum_{b=0}^{l_w-1} w_i^{(b)} \cdot \mathbb{B}_i[b]$, and each bit $w_i^{(b)}$ is multiplied with the activation with OT. The final results are acquired by combining the partial results by shift and addition.
Residual protocol follows Section \ref{sec:residual}.
\section{Experimental Results}

\begin{table}[!tb]
    \centering
    \caption{Micro-benchmark (MB) of convolution protocol. 
    % where $H, W, C, K$ denote the activation height, width, input channels, and output channels, respectively.
    }
    \label{tab:conv_micro_bench}
    \resizebox{\linewidth}{!}{
    \begin{tabular}{cccc|cccc}
        \bottomrule
         \textbf{Height} & \textbf{Width} & \tabincell{c}{\textbf{\#Input}\\\textbf{Channel}}  & \tabincell{c}{\textbf{\#Output}\\\textbf{Channel}} & \textbf{SiRNN}  &  \textbf{CoPriv}  &  \cellcolor{orange!10} \tabincell{c}{\textbf{\method}\\\textbf{(W2A4)}}  & \cellcolor{orange!10} \tabincell{c}{\textbf{\method}\\\textbf{(W2A6)}} \\
        \hline                         
        32 & 32 & 16 & 32 &  414.1  & 247.3    &    \cellcolor{orange!10} 25.75     &  \cellcolor{orange!10} 30.88    \\
        16& 16& 32& 64 &   380.1 & 247.3   &  \cellcolor{orange!10} 17.77     &  \cellcolor{orange!10} 20.33   \\
        56& 56& 64& 64& 9336 & 5554 & \cellcolor{orange!10} 376.9  & \cellcolor{orange!10} 440.6  \\
        28& 28& 128& 128& 8598 & 5491  & \cellcolor{orange!10} 381.9  & \cellcolor{orange!10} 438.9  \\
        \bottomrule
    \end{tabular}
    }
\end{table}

\begin{figure}[!tb]
    \centering
    \includegraphics[width=\linewidth]{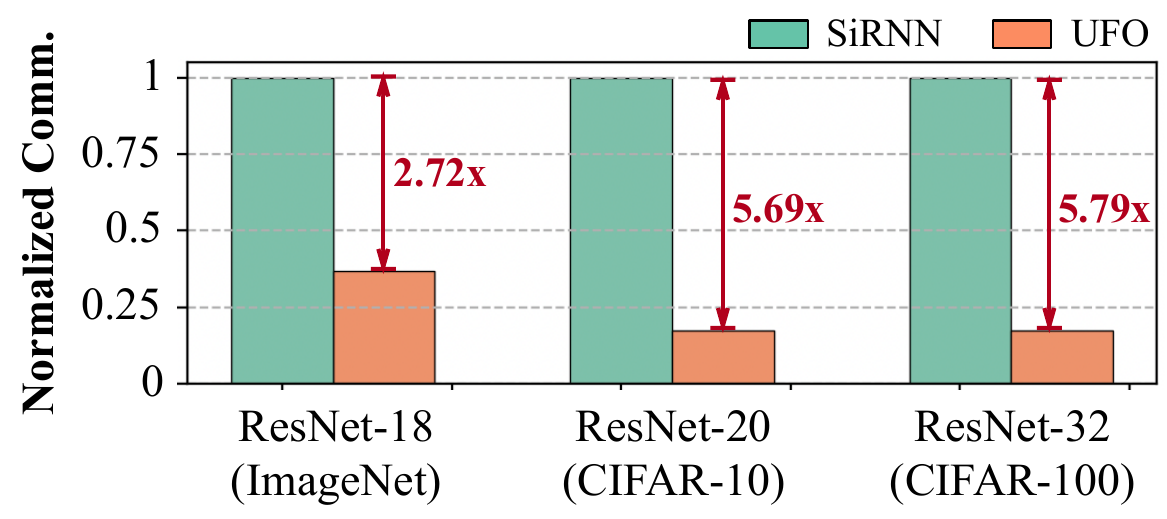}
    \caption{Micro-benchmark of residual protocol.}
    \label{tab:res_bench}
    \vspace{-10pt}
\end{figure}

\subsection{Experimental Setups}

\subsubsection{\textbf{Inference Framework}}
\method~is implemented on the top of SiRNN \cite{rathee2021sirnn} in EzPC\footnote{\url{https://github.com/mpc-msri/EzPC}} library in C++ for private inference.
Specifically, our quantized Winograd convolution is implemented in C++ with Eigen and Armadillo libraries.
% Note that homomorphic encryption (HE) based convolution achieves lower communication compared to OT-based convolution at the cost of more computation overhead for both the server and client. 
% Hence, we believe HE and OT have different applicable scenarios. 
% For example, for less performant clients, HE may not be applicable, while when the bandwidth is low, OT may not be the best choice.
% Note that \method~can be applied to both MPC and HE frameworks for better efficiency.
Following \cite{huang2022cheetah,rathee2020cryptflow2}, we use LAN mode for communication,
where the bandwidth is 377 MBps and echo latency is 0.3ms.
In our experiments, we evaluate the inference efficiency on the
Intel Xeon Gold 5220R CPU @ 2.20GHz.

% For quantization, we adopt QAT for the Winograd-based networks.

\subsubsection{\textbf{Models and Datasets}}
We conduct experiments on various models and datasets, i.e., MiniONN \cite{liu2017oblivious} and ResNet-20 on CIFAR-10, ResNet-32 on CIFAR-100, ResNet-18 on Tiny-ImageNet and ImageNet \cite{deng2009imagenet}.
Baseline DeepReDuce \cite{jha2023deepreshape}, SNL \cite{cho2022selective}, SAFENet \cite{lou2020safenet}, and SENet \cite{kundu2023learning} are ReLU-reduced methods evaluated using SiRNN.

\subsubsection{\textbf{QAT Details}}
To improve the performance of quantized models, we adopt quantization-aware training with the PyTorch framework for Winograd-based models.
% We finetune the quantized models using the PyTorch framework.
During training, we fix the transformation matrices $A, B$ while $G$ is set to be trainable since $GWG^\top$ is computed and quantized offline.
% For each model, we fix the bit width of the first and last layer to 8 bits. 
For MiniONN, we fix the bit width of activation to 4-bit while for ResNets, we fix it to 6-bit, and we only search the bit widths for weights.
We always use high precision, i.e., 8 bits, for the residual for better accuracy \cite{wu2018mixed,yang2021fracbits,xu2024privquant}.

\subsection{Micro-benchmark Evaluation}
\label{sec:micro}
% To verify the effectiveness of our proposed protocols, we compare \method~with SiRNN \cite{rathee2021sirnn} and CoPriv \cite{zeng2023copriv}.

\subsubsection{\textbf{Convolution Protocol}}
To verify the effectiveness of our proposed QWinConv protocol, we benchmark the convolution communication in Table \ref{tab:conv_micro_bench}.
We compare \method~with SiRNN \cite{rathee2021sirnn} and CoPriv \cite{zeng2023copriv} given different layer dimensions.
As observed, with W2A4,
\method~achieves 15.8$\sim$24.8$\times$ and 9.6$\sim$14.7$\times$ communication reduction, compared to SiRNN and CoPriv, respectively.

\subsubsection{\textbf{Residual Protocol}}
In Figure \ref{tab:res_bench}, we compare our simplified residual protocol with SiRNN.
We focus on the comparison with SiRNN as other works usually ignore the protocol and suffer from computation errors.
\method~achieves 2.7$\times$, 5.7$\times$ and 5.8$\times$ communication reduction on ResNet-18, ResNet-20, and ResNet-32, respectively.

\begin{figure}[!tb]
    \centering
    \includegraphics[width=\linewidth]{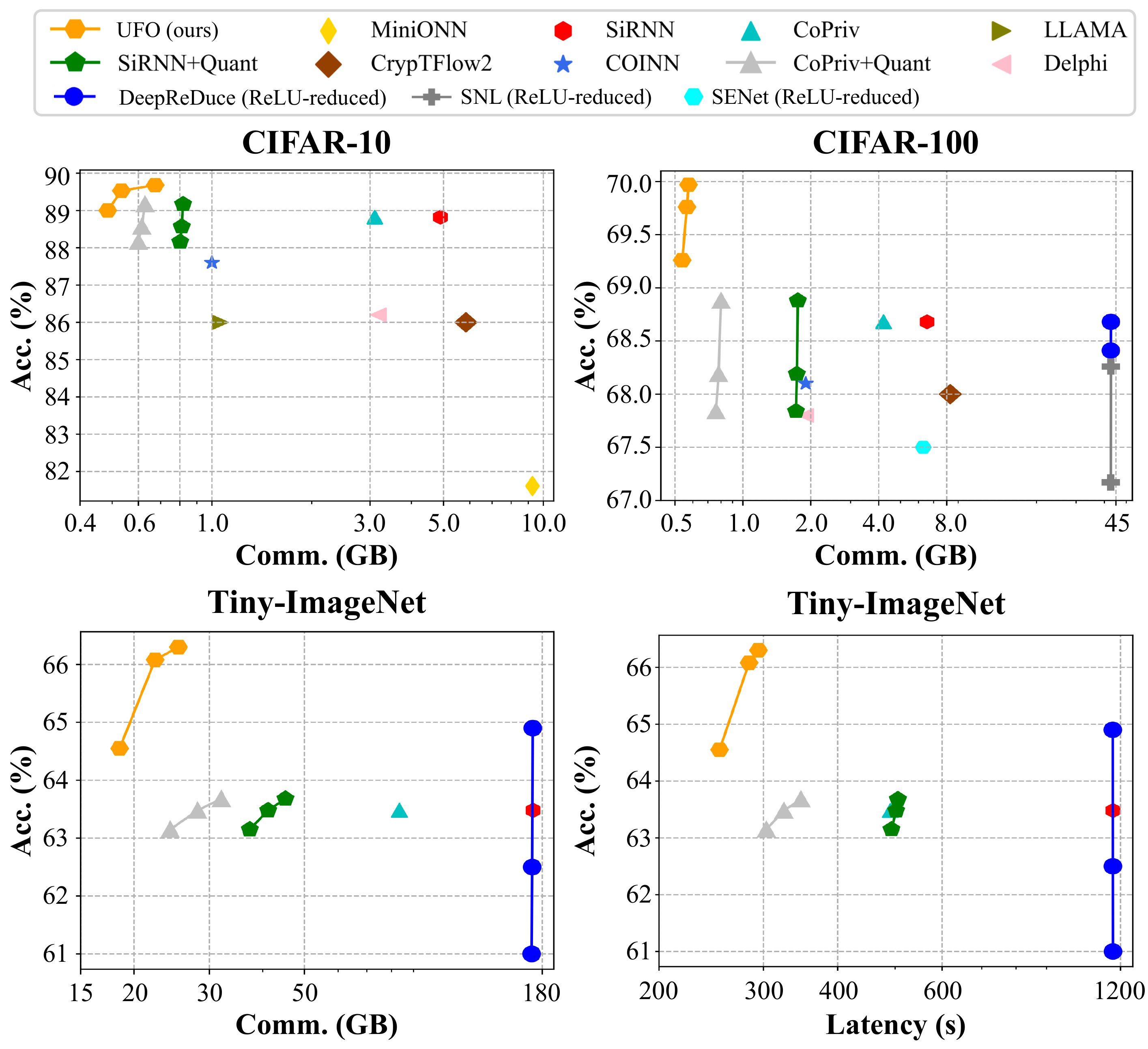}
    \caption{End-to-end comparison with prior-art methods.}
    \label{fig:main_result}
    \vspace{-15pt}
\end{figure}

\subsection{End-to-end Inference Evaluation}

\subsubsection{\textbf{CIFAR-10 and CIFAR-100}}
From Figure \ref{fig:main_result}, we make the following observations:
\underline{1)} \method~achieves state-of-the-art Pareto front of the accuracy and efficiency. 
Specifically, on CIFAR-10, \method~outperforms SiRNN with 0.71\% higher accuracy and 9.41$\times$ communication reduction.
On CIFAR-100, \method~achieves 1.29\% higher accuracy with 11.7$\times$ and 6.33$\times$ communication reduction compared with SiRNN and CoPriv, respectively;
\underline{2)} compared with COINN, \method~achieves 1.4\% and 1.16\% higher accuracy with 2.1$\times$ and 3.6$\times$ communication reduction on CIFAR-10 and CIFAR-100, respectively;
\underline{3)} when compared with ReLU-reduced methods,
including DeepReDuce \cite{jha2021deepreduce}, SAFENet \cite{lou2020safenet} and SNL \cite{cho2022selective}. 
The result shows these methods cannot effectively reduce total communication, and 
\method~achieves more than 80$\times$ and 15$\times$ communication reduction with higher accuracy, compared with DeepReDuce/SNL and SAFENet, respectively.
% \underline{4)} compared with a recent work LLAMA \cite{cryptoeprint:2022/793} on CIFAR-10, \method~achieves 2.24$\times$ communication redcution with 3\% higher accuracy.

\subsubsection{\textbf{Tiny-ImageNet and ImageNet}}
\begin{figure}[!tb]
    \centering
    \includegraphics[width=\linewidth]{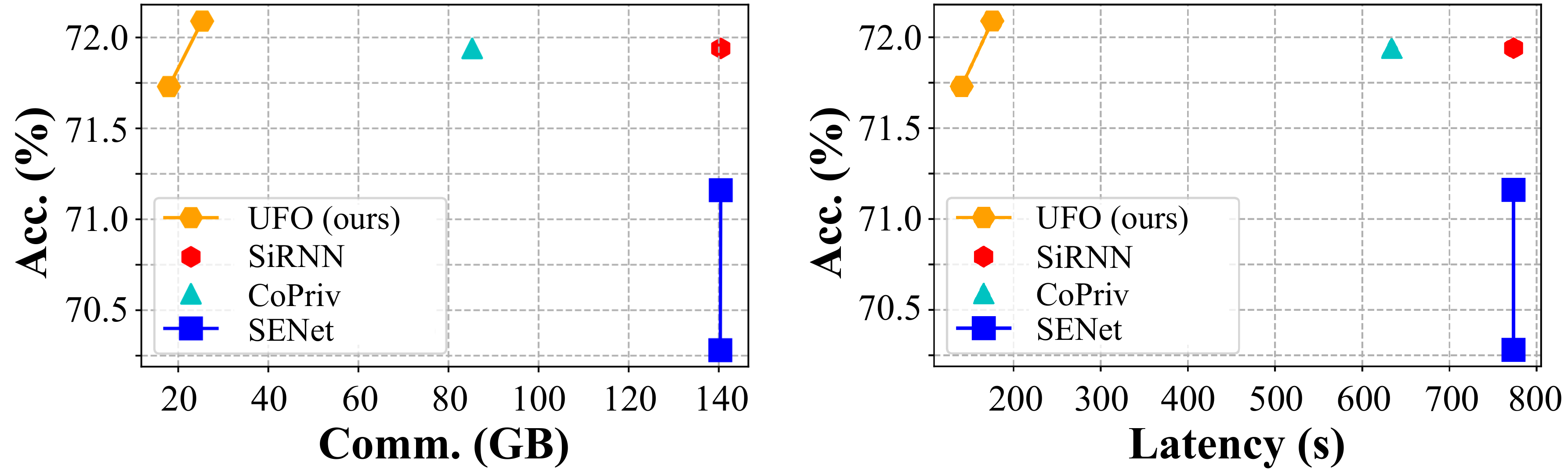}
    \caption{Evaluation on ImageNet.}
    \label{fig:imagenet}
\end{figure}

We compare \method~with SiRNN, CoPriv, and ReLU-optimized method DeepReDuce on Tiny-ImageNet.
As shown in Figure \ref{fig:main_result}, we observe that
\underline{1)} \method~achieves 9.26$\times$ communication reduction with 1.07\% higher accuracy compared with SiRNN;
\underline{2)} compared with SiRNN with mixed-precision, \method~achieves 2.44$\times$ communication reduction and 0.87\% higher accuracy;
\underline{3)} compared with CoPriv, \method~achieves 4.5$\times$ communication reduction with 1.07\% higher accuracy.
% \underline{4)} in terms of LAN latency evaluation, \method~reduces the SiRNN latency by 2.3$\times$, which demonstrates the consistency with communication reduction.
We also evaluate the accuracy and efficiency of \method~on ImageNet in Figure \ref{fig:imagenet}, \method~achieves 4.88$\times$ and 2.96$\times$ communication reduction with 0.15\% higher accuracy compared with SiRNN and CoPriv, respectively.

\subsection{Comparison with Prior-art Winograd Quantization}
We compare \method~with prior-art Winograd quantization algorithms, including Legendre \cite{barabasz2020quantaized}, BQW \cite{chikin2022channel}, and PAW \cite{chen2024towards} on ResNet-20 and CIFAR-10.
In Table \ref{tab:compare_quant},
\method~consistently achieves higher accuracy and higher inference efficiency, demonstrating the effectiveness of our quantization algorithm.
For example, compared to WinoVidiVici \cite{mori2024wino}, \method~achieves 2.63$\times$ communication and 4.37$\times$ latency reduction, respectively with 0.28\% higher accuracy.

\begin{table}[!tb]
    \centering
    \caption{Comparison with prior-art Winograd quantization methods on CIFAR-10.}
    \label{tab:compare_quant}
    \resizebox{\linewidth}{!}{
    \begin{tabular}{c|ccc}
    \bottomrule
    \textbf{Method}     &  \textbf{Comm. (GB)}&  \textbf{Lat. (s)} & \textbf{Top-1 Acc. (\%)}  \\
    \hline
    PTQ \cite{li2021brecq}                     & 0.783& 56.85  & 80.63   \\
    \hline
    Legendre \cite{barabasz2020quantaized}     & 0.965& 61.86  & 91.80   \\
    BQW \cite{chikin2022channel}               & 0.783& 56.16  & 90.74   \\
    PAW \cite{chen2024towards}                 & 0.784& 57.01  & 90.25   \\
    WinoVidiVici \cite{mori2024wino}           & 0.964& 60.40  & 92.29   \\
    \rowcolor{orange!10}
    \method~(ours)                             & 0.366& 13.82  & 92.57   \\
    \bottomrule
    \end{tabular}
    }
\end{table}

\vspace{-5pt}
\subsection{Comparison with HE-based Frameworks}
We also compare \method~with representative HE-based methods Gazelle \cite{juvekar2018gazelle}, Delphi \cite{mishra2020delphi}, CrypTFlow2 \cite{rathee2020cryptflow2} on CIFAR-100 in Table \ref{tab:he}, and \method~achieves significantly higher efficiency and accuracy. 
For example, compared to CrypTFlow2 \cite{rathee2020cryptflow2}, \method~achieves 1.30$\times$ and 4.87$\times$ communication and latency reduction, respectively.
% compared with Gazelle, Delphi, and CrypTFlow2.
% Compared with SOTA Cheetah, although \method~has higher communication, it still achieves 1.7$\times$ lower latency.
% We believe \method~provides a promising solution for the MPC research community.
% empirically demonstrating the feasibility of \method.
Additionally, the proposed \method~can be generally applied to HE-based frameworks and protocols, which we leave to our future work.

\begin{table}[!tb]
    \centering
    \caption{End-to-end comparison with HE-based methods.}
    \label{tab:he}
    \resizebox{\linewidth}{!}{
    \begin{threeparttable}
    \begin{tabular}{c|ccc}
    \bottomrule
    \textbf{Method}    &  \textbf{Comm. (GB)} & \textbf{Lat. (s)} &  \textbf{Top-1 Acc. (\%)}  \\
    \hline
    Gazelle \cite{juvekar2018gazelle}    &  8.900 &  238.7 & 67.90  \\
    Delphi \cite{mishra2020delphi}     &  6.500   & 200.0 & 67.80   \\ 
    % Cheetah  &  0.202  &  34.87 &  68.00   \\
    CrypTFlow2 \cite{rathee2020cryptflow2}\tnote{1} &   0.681   &  99.03  & 68.00  \\
    \rowcolor{orange!10}
    \method~(ours)  &  0.525 & 20.33 & 69.26  \\
    \bottomrule
    \end{tabular}
    \begin{tablenotes}
        \scriptsize
        % \item[1]The protocol of the pre-processing phase is HE-based.
        \item[1]Use the version that linear layers are implemented using HE protocol.
      \end{tablenotes}
    \end{threeparttable}
    }
    % \vspace{-10pt}
\end{table}

\subsection{Ablation Studies}

\begin{figure}[!tb]
    \centering
    \includegraphics[width=\linewidth]{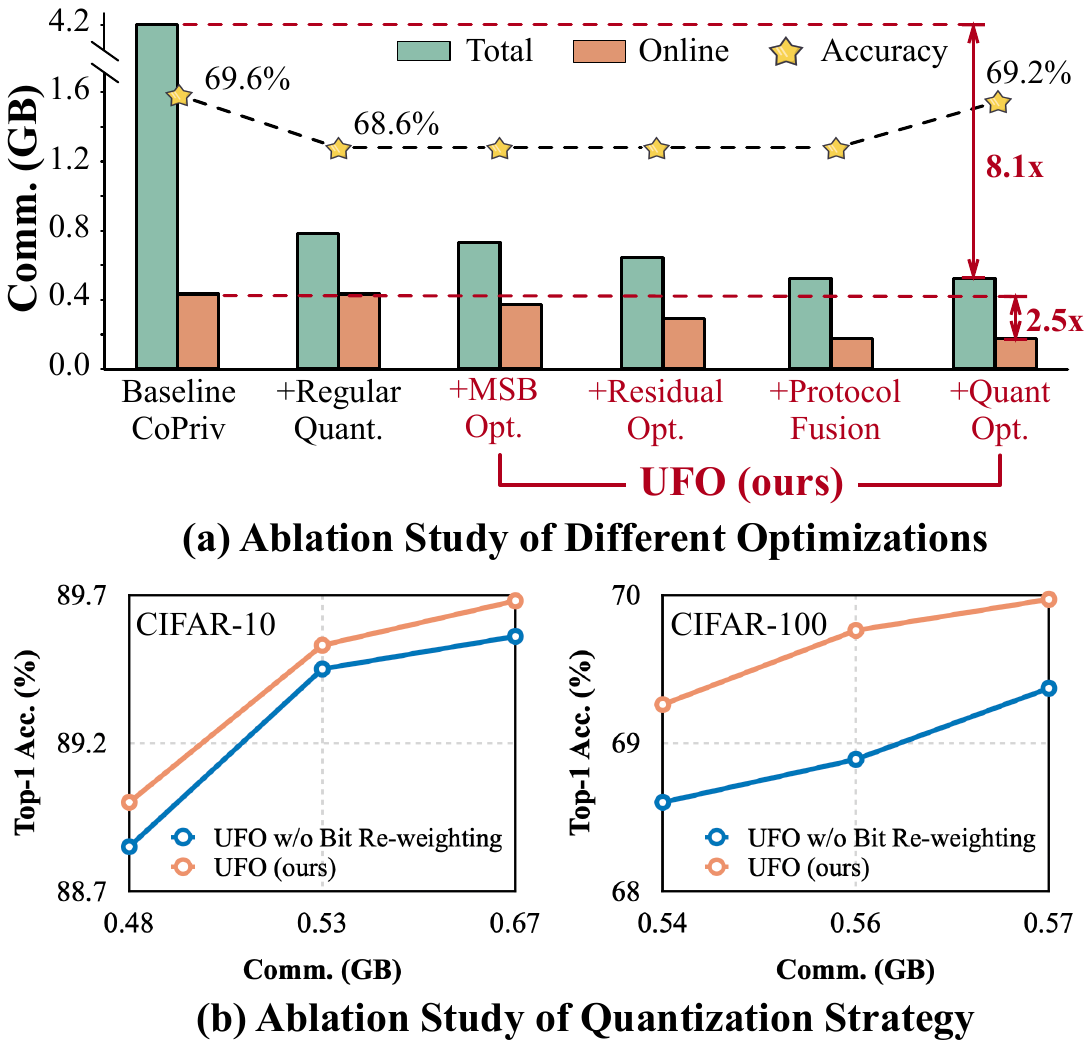}
    \caption{Effectiveness of our proposed optimizations.}
    \label{fig:ablation}
    \vspace{-10pt}
\end{figure}

\subsubsection{\textbf{Effectiveness of Different Optimizations}}
To understand how different optimizations help improve 2PC inference, we add the optimizations step by step on ResNet-32 and present the results in Figure \ref{fig:ablation}.
As observed from the results, we find that
\underline{1)} low-precision quantization benefits the total communication efficiency most but there is no benefit for online communication due to the extra bit width conversions introduced from Winograd transformations;
\underline{2)} protocol optimizations including the simplified residual protocol, protocol fusion, and MSB-known optimization 
consistently reduce the online and total communication,
\underline{3)} although naively combining Winograd convolution with quantized private inference enlarges the online communication, our optimizations jointly achieve 8.1$\times$ and 2.5$\times$ total and online communication reduction, respectively, compared to CoPriv;
\underline{4)} bit re-weighting improves the accuracy without sacrificing efficiency.
The findings indicate all of our optimizations are indispensable for 2PC inference.

% \textbf{Effectiveness of Quantization Strategy.}
% As shown in Figure \ref{fig:ablation}(b), our proposed quantization algorithm achieves 0.1$\sim$0.2\% higher accuracy and more than 0.6\% higher accuracy on CIFAR-10 and CIFAR-100 without sacrificing efficiency, respectively, demonstrating the efficacy of \method.
% Since we re-weight $2^{n-1}$ to $2^n$ with 1-bit increment and there are around 20\% layers that have large outliers, the communication is only slightly affected.

\subsubsection{\textbf{Block-wise Visualization}}
\begin{figure}[!tb]
    \centering
    \includegraphics[width=\linewidth]{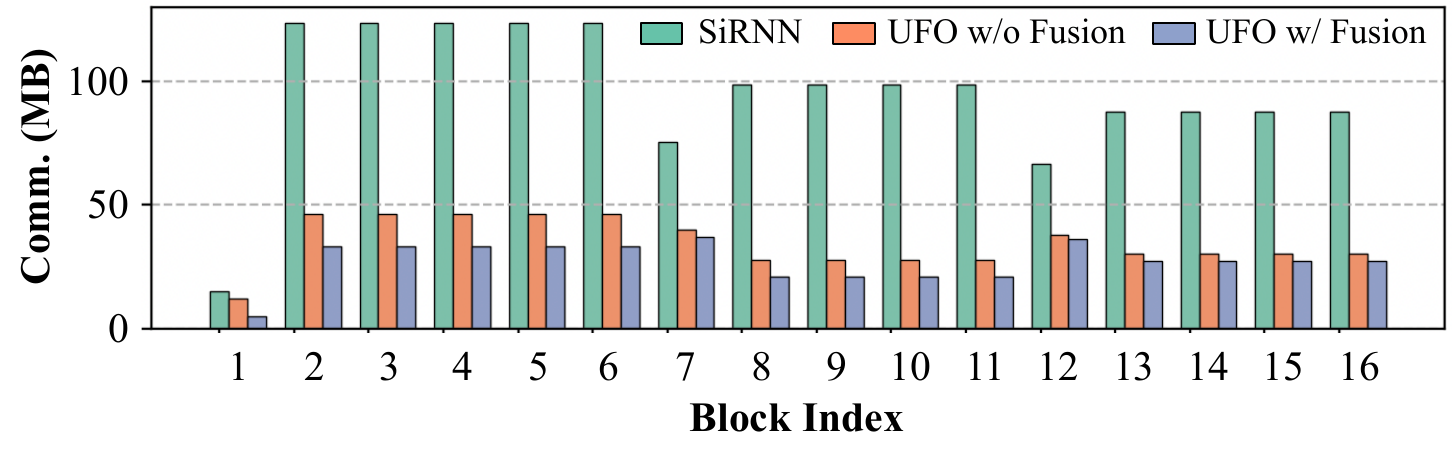}
    \caption{Block-wise visualization of communication.}
    \label{fig:abl_fuse}
    % \vspace{-10pt}
\end{figure}

% To demonstrate the effectiveness of protocol fusion, we take W2A6 as an example on ResNet-32 and CIFAR-100.
We show the block-wise communication comparison on ResNet-32
as shown in Figure \ref{fig:abl_fuse}.
As observed,
with our quantized Winograd convolution protocol, \method~achieves 63\% communication reduction on average, and 
% 2.7$\times$
% To further leverage protocol fusion, \method~reduces 3.4$\times$ communication compared with SiRNN.
protocol fusion further reduces 30\% communication.
The results demonstrate the effectiveness of \method~and different layers benefit
from \method~differently.
% We also find the communication portion of bit width conversions for Winograd transformation in the early layer is larger than in the later layers.

% \begin{figure}
%     \centering
%     \includegraphics[width=0.6\linewidth]{figure/ablation_fuse.pdf}
%     \caption{Ablation study and block-wise visualization of protocol fusion for the extension of Winograd transformation.}
%     \label{fig:abl_fuse}
%     % \vspace{-14pt}
% \end{figure}

\subsubsection{\textbf{Distribution of Bit Width Assignment}}
\begin{figure}
    \centering
    \includegraphics[width=\linewidth]{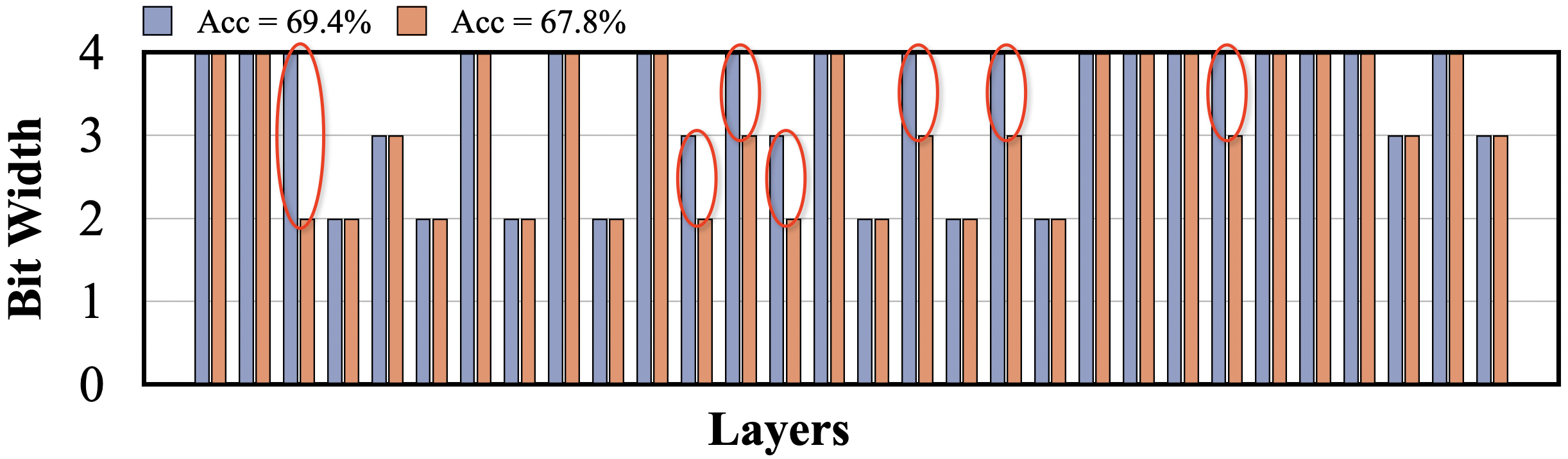}
    \caption{Distribution of bit width assignment on ResNet-32.}
    \label{fig:distribution_bw}
    \vspace{-10pt}
\end{figure}

We visualize the distribution of bit width assignment by the sensitivity-based method in Figure \ref{fig:distribution_bw}.
As observed, the model tends to use lower bit widths in the middle layers rather than the initial and last layers.
Meanwhile, a lower bit width allocation often corresponds to lower bit widths in the middle layers.

\section{Related Work}
\label{sec:related}
% \wx{@Chao Yang: Extend this paragraph (refer to our survey)}

There has been an increasing amount of literature on OT-based private inference, 
which can be generally categorized into the following three paradigms.
% including protocol optimization,
% \cite{demmler2015aby,rathee2020cryptflow2,rathee2021sirnn,knott2021crypten,mohassel2017secureml,xu2025breaking,rathee2022secfloat,lu2023bumblebee,pang2023bolt,hao2022iron}
% algorithm optimization, and 
% \cite{cho2022selective,kundu2023learning,li2022mpcformer,zeng2023mpcvit,rathee2024mpc,zeng2025mpcache,wu2024ditto,peng2023autorep,li2024seesaw,cai2022hunter,xia2025cryptpeft,chen2022x}
% algorithm-protocol co-optimization.
% \cite{mishra2020delphi,hussain2021coinn,zeng2023copriv,xu2024privquant,xu2024privcirnet}.

\subsubsection{\textbf{Protocol Optimization}}
A substantial body of work, such as
CryptFlow2 \cite{rathee2020cryptflow2}, CrypTen \cite{knott2021crypten}, SecureML \cite{mohassel2017secureml} leverage OT to evaluate both linear and non-linear layers.
SiRNN \cite{rathee2021sirnn} further extends the protocols to mixed bit widths, while Secfloat \cite{rathee2022secfloat} extends the protocols to floating-point representation.
Squirrel \cite{lu2023squirrel} builds new protocols to support gradient boosting decision tree (GBDT) training.
CipherGPT \cite{hou2023ciphergpt} optimizes the GEMM using VOLE to significantly reduce communication overhead.
Studies like \cite{lu2023bumblebee,pang2023bolt,hao2022iron,xu2025breaking} use OT to evaluate the non-linear layers.

\subsubsection{\textbf{Algorithm Optimization}}
Many research efforts have been made to linear layer optimization \cite{ganesan2022efficient,cai2022hunter}, ReLU and GeLU optimization \cite{cho2022selective,kundu2023learning,peng2023autorep,zeng2023mpcvit,jha2021deepreduce,li2024seesaw}, and Softmax optimization \cite{zeng2023mpcvit,zeng2025mpcache,li2022mpcformer,dhyani2023privit}.
Classically, MPCFormer \cite{li2022mpcformer} replaces Softmax with MPC-friendly polynomial approximations.
% and employs knowledge distillation to mitigate accuracy degradation. 
MPCViT \cite{zeng2023mpcvit} replaces the exponential in Softmax with ReLU and selectively uses linear attention using neural architecture search (NAS).
CryptPEFT \cite{xia2025cryptpeft} confines expensive cryptographic operators within lightweight adapters and utilizes NAS to identify optimal adapter architecture.

\subsubsection{\textbf{Co-optimization}}
Besides optimizing the protocol and algorithm alone,
Delphi \cite{mishra2020delphi} transfers most of the online cost to the offline phase and uses NAS to reduce ReLUs. 
COINN \cite{hussain2021coinn} proposes factorized GEMM by weight clustering and a corresponding protocol to reduce the number of multiplications.
% and evaluate non-linear layers using GC.
CoPriv \cite{zeng2023copriv} integrates the Winograd convolution protocol, ReLU pruning, and layer fusion to cut down the overall communication cost.
PrivQuant \cite{xu2024privquant} uses mixed-precision quantization with protocol fusion and MSB-known optimization to reduce the overhead.

Another line of work attempts to reduce inference overhead with quantization \cite{samragh2021application,hussain2021coinn,rathee2021sirnn,shen2022abnn2,hasler2023overdrive,agrawal2019quotient,wu2024ditto}.
For instance, Ditto \cite{wu2024ditto} adopts static dyadic quantization to avoid dynamically computing scale during inference and proposes type conversion protocols for efficient bit width conversion.
In Table \ref{tab:compare_exist}, we compare \method~with previous works qualitatively. 
% As observed, \method~jointly optimizes both protocol and algorithm to simultaneously reduce the number of multiplication and communication per multiplication for efficient 2PC inference.
\section{Conclusion and Future Work}

% \textbf{Limitation.}

% \textbf{Conclusion.} 

In this work, we propose \method, an efficient 2PC-based framework seamlessly combining Winograd convolution and quantization.
% We observe the bit width and the number of multiplications jointly determine the total communication.
Based on the observation that 
naively combining quantization and Winograd is sub-optimal,
% We first analyze the principle of the OT protocol, and find both the bit width and the number of multiplications are important to total communication.
% To make quantization compatible with Winograd,
we propose a series of protocol optimizations to minimize the communication
and develop a mixed-precision QAT algorithm with a bit re-weighting algorithm to improve the accuracy and efficiently accommodate quantization outliers.
% we propose to combine Winograd convolution and quantized 2PC-based inference together along with a series of graph-level optimizations to further reduce communication.
Extensive experiments demonstrate that \method~consistently enhances the efficiency without compromising the accuracy compared with the prior-art baselines across a variety of protocols and algorithms.

This work utilizes the Winograd algorithm to optimize the convolution layers, which is tailored for CNNs.
% Since the Winograd algorithm cannot be directly used for Transformer-based architectures, we do not focus on Transformers in this paper, and we leave it as our future research.
Given that Winograd-based acceleration does not directly translate to Transformer-based architectures, the scope of this paper is focused on CNNs. Extending our methodology to Transformers remains a promising direction for future research.
Moreover, our future work also includes investigating Winograd optimizations with HE protocols. This necessitates specialized designs for HE-friendly encoding and quantization algorithms.

% \clearpage
\bibliographystyle{IEEEtran}
\bibliography{main}

\end{document}